\documentclass{aa}
\usepackage[varg]{txfonts}
\usepackage{graphicx}
\usepackage{natbib}
\usepackage{textcomp}
\usepackage{gensymb}
\usepackage{xcolor}
\usepackage{hyperref}
\usepackage{amsmath}
\usepackage{array}
\usepackage{tabularx}
\bibpunct{(}{)}{;}{a}{}{,}
\usepackage{subcaption}
\usepackage{longtable}

\begin{document}

\title{COUGS-DESI: A Catalog of Unusual Galaxies with polar Structures in the DESI Legacy Imaging Surveys}

\author{Seneca K. H. Bahr\inst{1} 
\and Aleksandr V. Mosenkov\inst{1} 
\and Jacob A. Guerrette\inst{1}
\and Isaac H. Jensen\inst{1}
\and Jonah X. George\inst{2} 
\and Thea~E.~Spigarelli\inst{1}
\and Ryan P. Smith\inst{3}
\and Brandon T. Burton\inst{1}
\and Kevan W. Beckstead\inst{1}
\and Jonah D. Seguine\inst{1}
\and Harrison K. Casper\inst{1}
\and Nefi Pineda\inst{1}
\and Michael Holland\inst{1}}

\institute{Department of Physics and Astronomy, N283 ESC, Brigham Young University, Provo, UT 84602, USA
\and Department of Astronomy, University of Maryland, College Park, MD 20742, USA
\and Department of Physics, University of Illinois at Urbana-Champaign, Urbana, IL 61801, USA}

\date{Received 12 December 2025 / Accepted 7 April 2026}

\abstract{
\textit{Context.} Polar-structure galaxies (PSGs) host photometrically and kinematically decoupled components oriented at large angles to one another. These systems, which include polar rings, polar disks, polar halos, polar bulges, polar dust lanes, and polar tidal structures, provide valuable insights into galaxy formation and evolution, although statistical studies are limited by their rarity.

\textit{Aims.} We aim to construct the largest and most homogeneous catalog of PSGs to date to enable robust statistical studies of their properties and occurrence rates in the Local Universe.

\textit{Methods.} Using DESI Legacy Imaging Surveys (DR10) data, we identified PSG candidates in the Siena Galaxy Atlas (SGA) through visual inspection, convolutional neural network (CNN) classification, and cross-matching with previously reported systems. Each galaxy was assigned a PSG subtype and host morphology. We analyzed the general properties of PSGs and compared them with those for all galaxies in the SGA. Simple image simulations were used to evaluate projection effects.

\textit{Results.} The resulting Catalog of Unusual Galaxies with polar Structures in the DESI Legacy Imaging Surveys (COUGS-DESI) contains 2,989 PSG candidates, including 342 previously known objects. Projection effects from random galaxy overlaps are negligible. The sample spans a wide range of polar morphologies, with 1,113 polar rings, 75 polar bulges, 216 polar halos, 185 polar dust lanes, and 1,315 polar tidal structures. Overall, PSGs constitute $2.2\%$ of local non-dwarf galaxies, with polar rings representing $0.7\%$.  Approximately 1\% of S0 galaxies in the SGA host polar rings, whereas spirals constitute the most common morphological type among the PSG hosts in our catalog.

\textit{Conclusions.} COUGS--DESI increases the number of known PSG candidates by an order of magnitude and provides a foundation for detailed studies of the formation and evolution of polar structures.
}


\titlerunning{COUGS-DESI: Polar-structure galaxies in DESI}
\authorrunning{Bahr et al.}

\maketitle
\nolinenumbers

\section{Introduction} \label{sec:intro}

The Universe features a remarkable diversity that goes beyond the familiar classes of regular galaxies, whereby more exotic objects reveal structures that testify to the complexity of galaxy formation and evolution. Among them, polar-structure galaxies (PSGs) represent a special case --- not only are they visually striking, but they also serve as unique probes of galaxy formation and evolution. They exhibit prominent large-scale structural features oriented at a significant angle to the principal plane of the host galaxy. Their unique structural configurations, combined with their histories of past or ongoing interaction with other galaxies and their cosmological environment, make them excellent tools for addressing the question of how galaxies form and evolve. The photometric and kinematic decoupling of the components in PSGs also makes them ideal natural laboratories for studying the interplay of complex physical processes, including gas accretion \citep{2008ApJ...689..678B,2014A&A...569A..83C}, star formation \citep{2015MNRAS.447.2287R,2025PASA...42...56A}, active galactic nucleus (AGN) activity \citep{2020AstL...46..501S,2022MNRAS.512.2556H}, and both major and minor mergers \citep{1997A&A...325..933R,2003ApJ...587L..23S,2003A&A...401..817B,2016A&A...585A.156O,2020MNRAS.491.1887Q,2022RAA....22k5003M}. Such processes are intricately linked together within the galactic environment (see e.g.,\citealt{2006PASP..118..517B}), and are also influenced by the dark matter distribution (see a review by \citealt{2018ARA&A..56..435W}). Therefore, studying PSGs holistically is both necessary and important for our general understanding of galaxy assembly and dynamics.

Under the hierarchical model of galaxy evolution within the lambda cold dark matter ($\Lambda$CDM) paradigm of galaxy formation and evolution, galaxies evolve over time through gravitational interactions, including mergers \citep[see e.g.][]{2020NatRP...2...42V}. Such interactions leave observable signatures in the form of a variety of stellar streams and tidal tails. As a result, massive galaxies in the Local Universe are expected to have transient tidal features arising from minor mergers \citep{2010AJ....140..962M, 2023A&A...671A.141M, 2025A&A...700A.176M} positioned at a variety of angles, many of which may have orthogonal orientations and lifetimes up to a few Gyr before they broaden significantly or fade below detection thresholds \citep{2025A&A...700A.176M}. Such structures can be classified as polar tidal structures and, in some cases, may contribute to the formation of other, more long-lived and well-defined polar structures.

Polar-ring galaxies (PRGs), a well-studied subtype of PSGs consisting of a central object (host galaxy) orbited by a ring or disk of stars and gas in a nearly orthogonal plane, are differentiated from polar tidal structures by the longevity of their polar components. While still transient, some polar structures in semi-stable configurations may endure for several to more than $8$~Gyr, depending on the configuration of the dark matter halo and other processes \citep{1990AJ....100.1489W,1993AJ....105.1378P,2003A&A...401..817B,2025PASA...42...56A}. Furthermore, due to perturbations from the host galaxy, polar rings and disks can develop spiral arms over their lifetimes \citep{1997PASA...14...92A, 2006A&A...446..905T}. 

The concept of PSGs as a broader structural family extending beyond PRGs to include all galaxies with significant orthogonal or nearly orthogonal features relative to the major axis of the host galaxy was proposed by \citet{2024A&A...681L..15M}. PSGs exhibit a variety of structures with different physical natures, but they may appear similar in images due to, for example, poor angular resolution and specific viewing angles of the host and polar structure. For instance, blue, disrupted satellites on radial orbits in highly inclined galaxies can resemble edge-on polar rings in optical images, as can polar optical jets associated with AGNs. Furthermore, the different types of polar structures may show significant overlap in their formation mechanisms, as demonstrated by observational data and cosmological simulations \citep{1997ApJ...490L..37B, 1998ApJ...499..635B, 2003A&A...401..817B, 2004AJ....128..137M, 2008ApJ...689..678B, 2012MNRAS.423L..79C, 2019MNRAS.485..464L, 2021MNRAS.504.5702W, 2023MNRAS.519.4735S}. Since PSGs appear morphologically similar and have similar formation mechanisms, it is likely that they are related despite their diversity \citep{2024A&A...681L..15M}. We therefore chose to treat PSGs as a unified broad class of objects in this study. 

This proposed broader class of PSGs includes PRGs, polar bulges \citep[][]{2012MNRAS.423L..79C,2015AstL...41..748R}, polar halos \citep[][]{2020MNRAS.494.1751M,2022RAA....22k5003M,2022MNRAS.512.2556H}, polar dust lanes \citep[which often correspond to gaseous polar rings;][]{1978ApJ...226L.115B,2009MNRAS.393..317S}, and polar tidal structures \citep{2008ApJ...689..184M,2023A&A...671A.141M}. Taken together, PSGs represent a significant yet underexplored class of objects, united by the presence of visibly detected polar structures. The goal of unifying these diverse subtypes under the PSG framework is to provide a common observational basis for their identification. In some cases, photometric data alone are insufficient to distinguish among these subtypes, and there is a need for spectroscopic follow-up to disentangle their true physical nature.

The astronomical community began to take an interest in PRGs and other PSGs in the 1960s and 1970s \citep{1961hag..book.....S, 1978ApJ...226L.115B, 1978AJ.....83.1360S, 1983AJ.....88..909S}. Traditionally, studies of PSGs have been limited to individual cases \citep{1978AJ.....83.1360S, 1995AJ....109..942V, 2006A&A...451...99B, 2013A&A...560A..14P, 2015MNRAS.450..998S, 2020MNRAS.491.1887Q, 2022PASP..134i4105N} or small samples \citep{2012AstBu..67..147M, 2012MNRAS.422.2386F, 2013AstBu..68..371S, 2019MNRAS.486.4186E, 2022RAA....22k5003M}. Although several catalogs of PRG candidates have been compiled in recent years \citep{1990AJ....100.1489W, 2011MNRAS.418..244M, 2019MNRAS.483.1470R}, the segregation between different PSG types (see Sect.~\ref{sec:cat_comp}) has made it challenging to generalize PSG properties and assess broader implications. Furthermore, only a small portion of the PSG candidates discussed in the literature have been kinematically confirmed (see e.g., \citealt{1983AJ.....88..909S,1990AJ....100.1489W, 2011MNRAS.418..244M, 2014ASPC..486...61M}; a complete summary is given in \citealt{2026ApJ...999..199Y}).

Part of the reason for the relatively small sample of genuine PRG candidates is the rarity of these objects and the difficulty related to their visual identification. Previously, the fraction of PRGs was estimated to be about 0.05--0.1\% of luminous galaxies in the Local Universe \citep{1990AJ....100.1489W, 2011AstL...37..171R, 2022MNRAS.516.3692S}. However, our recent pilot study \citep{2024A&A...681L..15M}, using deep Sloan Digital Sky Survey (SDSS, \citealt{2000AJ....120.1579Y}) Stripe\,82 data across 275 square degrees, identified 102 PSGs (including 53 PRGs), suggesting that such systems might constitute up to 1.1\% of luminous non-dwarf galaxies in the Local Universe. This fraction further rises as high as 3\% when inclination effects are taken into account and even higher if purely gaseous polar structures are included \citep{2023MNRAS.525.4663D}. In addition, the volume density of PRGs is expected to increase with redshift \citep{reshetnikov1997,2022MNRAS.516.3692S}. These findings suggest a significantly higher prevalence of PSGs than previously recognized. Although this is still a small fraction, it is much higher than previously assumed and should not be overlooked in galaxy morphology studies, particularly for early-type galaxies, where PSGs are more prevalent than in late types \citep{1990AJ....100.1489W}. 

Recent advances in deep optical imaging and observational techniques \citep[see e.g.,][]{2017ASSL..434..255K,2019MNRAS.490.1539R} have substantially mitigated previous observational limitations, creating the opportunity to identify a much larger sample of PSGs than was previously possible. With detection limits for deep photometric data from wide sky surveys such as the DESI Legacy Imaging Surveys (\citealt{2019AJ....157..168D}, DESI Legacy hereafter) reaching surface brightnesses as faint as $\mu_{g}\sim29 ~\mathrm{mag}\,\mathrm{arcsec}^{-2}$, it is feasible to identify a significant number of faint or distant PSG candidates across vast sky areas. The availability of deep wide-field photometry has enabled large-scale searches for, and studies of, faint extragalactic structures \citep{2022MNRAS.515.5335L, 2023A&A...671A.141M}. The pilot study by \citet{2024A&A...681L..15M} and other recent studies leveraging modern deep optical observations \citep{2019MNRAS.490.1539R, 2020MNRAS.494.1751M, 2020MNRAS.497.2039M, 2021MNRAS.503.6059P, 2023MNRAS.519.4735S, 2023MNRAS.525.3016M, 2024MNRAS.52710615M, 2024MNRAS.532..883S}, highlight the potential of deep photometry as a tool for expanding the known sample of PSGs.

In this article, using data from the DESI Legacy~DR10, we present the Catalog of Unusual Galaxies with polar Structures in the DESI Legacy Imaging Surveys (COUGS-DESI), a new catalog of PSGs that expands the existing sample size by an order of magnitude, constituting the largest PSG catalog to date. We establish a consistent nomenclature for PSGs, grounded in definitions commonly used in the literature, but refined to address important gaps. As the first comprehensive atlas encompassing all PSG subtypes, this catalog highlights the structural diversity of the PSG family and reveals such systems across the 20,000~deg$^{2}$ footprint of DESI Legacy, primarily using data from the Siena Galaxy Atlas \citep{2023ApJS..269....3M}. Furthermore, we provide the first large-scale statistical study of PSGs, drawing novel connections among the quasi-stable polar-structure types, including polar rings, polar bulges, and polar halos.

This paper introduces the catalog and atlas of PSGs. It also represents the first in a forthcoming series of studies dedicated to a detailed investigation of the PSG sample presented herein. In this paper, we describe and discuss the general properties of the selected PSGs in our catalog, as well as their occurrence rates in the Local Universe. Planned future works include a comparison of our observed PSG catalog with simulated PSGs from cosmological hydrodynamical simulations, an analysis of low-surface-brightness (LSB) features in PSGs to gain insights into their formation, and a photometric decomposition of PSGs with diverse structural components. We will also perform kinematic decomposition for a representative subset of PSGs and investigate their evolution with redshift within a cosmological framework. Finally, we plan to study AGN activity and star formation in the well-defined PSG sample, combining observational data with results from the simulations.

This paper is organized as follows. Section~\ref{sec:cat_comp} describes the methodology used to compile the catalog. Section~\ref{sec:cat_descr} presents a brief overview of the catalog and accompanying atlas of PSGs, together with the general statistics of the sample. In Section~\ref{sec:discussion}, we discuss the reliability of our classifications, the occurrence rates of PSGs and their subtypes, and factors that may affect these results, as well as the broader implications of this study. Finally, Section~\ref{sec:summary} summarizes our main findings and conclusions.

\section{Creation of the catalog}\label{sec:cat_comp}

Our PSG classification and analysis is based on imaging data from the DESI Legacy~DR10, which reaches an average depth of 28.5--29 mag arcsec$^{-2}$ (measured within a 10\textquotesingle\textquotesingle$\times$10\textquotesingle\textquotesingle\ box at the $3\sigma$ level in the $g$ band). This depth makes the photometry well suited for detecting and characterizing faint polar structures with average surface brightnesses down to 26--27 mag arcsec$^{-2}$ \citep{2024MNRAS.532..883S}. In this study, we have employed a dedicated method for generating enhanced RGB images from DESI Legacy data, designed to simultaneously highlight the inner and outer regions of galaxies---an essential step for the reliable identification of PSGs (see Appendix~\ref{app:rgb_method} for a detailed description of the method).

We began our search by examining PSG candidates reported in the literature, including both cataloged samples \citep{1990AJ....100.1489W, 2011MNRAS.418..244M, 2019MNRAS.483.1470R, 2024A&A...681L..15M} and individual case studies (primarily PRGs). Subsequently, we employed the Siena Galaxy Atlas 2020 \citep[SGA hereafter;][]{2023ApJS..269....3M} for the visual selection. The SGA covers approximately 20,000 deg$^2$ and contains $\sim$383,000 nearby ($z\lesssim0.3$) galaxies with sufficient angular diameters ($D(25) > 20$\textquotesingle\textquotesingle) for detailed morphological analysis \citep{2019MNRAS.483.1470R}. During our inspection of the DESI Legacy images for all SGA galaxies, we serendipitously identified an additional 65 PSGs with angular sizes too small to be included in the SGA, yet exhibiting clear polar structures; these were likewise incorporated into our catalog. In addition to the SGA, we also searched two catalogs of edge-on galaxies \citep{2014ApJ...787...24B, 2022MNRAS.511.3063M} and a new catalog of ringed galaxies for PSG candidates \citep{2025ApJS..280...11C}. To assist with the selection process, we also developed dedicated convolutional neural networks (CNN) optimized for PSG identification. The full catalog-construction procedure is described in the following sections. We begin, however, by providing a clear definition of a PSG and introducing the nomenclature of PSG subtypes used throughout this paper.

\subsection{PSG nomenclature}\label{subsec:nom}

We define a PSG as a galaxy that exhibits one or more visible large-scale structural components (i.e., stellar or dusty) whose projected major axis (associated with the angular momentum vector of the polar structure) is significantly misaligned, typically by $>40^\circ$ \citep{2013AstBu..68..371S}, with respect to the major axis of the main stellar disk or spheroid of the host galaxy. Although we use the term "polar," the more precise expression for these structures would be "polar and tilted structures;" however, for brevity, we use "polar" throughout this paper. Such components may include polar rings and polar disks (combined under the term PRGs), polar halos, polar bulges, or other extended features oriented in planes that are significantly tilted with respect to the host’s major axis (associated with the primary axis of rotation). The presence of these misaligned structures reflects a complex dynamical and evolutionary history involving external accretion, mergers, and internal dynamical instabilities.

In this paper, we provide a qualitative classification of these PSG types; a quantitative morphological and kinematic characterization will be presented in a subsequent study. For the purposes of this paper, we have only considered large-scale polar structures that are clearly identifiable in DESI Legacy images. Inner or small-scale polar features that are unresolved or invisible in regular DESI Legacy imaging were not included in the study. 

In brief, polar rings (PRs) are complete, closed rings or disks (sometimes exhibiting spiral arms, \citealt{1997PASA...14...92A,2006A&A...446..905T}) composed of material that may be predominantly stellar or gaseous \citep{1983AJ.....88..909S,1990AJ....100.1489W}. They are symmetrically centered on a host galaxy (typically a disk or elliptical) and oriented at a large angle (often close to 90$^{\circ}$) with respect to the major axis of the host. This category also includes edge-on polar rings, provided that the polar structure is relatively bright, symmetrical and centered on the host galaxy. It is believed that PRs arise from major or minor mergers, tidal accretion, or cold filamentary accretion (see Sect.~\ref{sec:intro}).

Polar bulges (PBs) are small, faint, prolate central components in disk galaxies \citep{2012MNRAS.423L..79C,2015AstL...41..748R}. PBs are often observed as secondary central structures that coexist with a classical bulge or a pseudobulge. Because of their small size and low luminosity, they are difficult to detect in face-on or moderately inclined systems using photometry alone. As a result, PBs are typically identified in edge-on galaxies. Disk galaxies with polar bulges are distinguished from polar-disk galaxies (in which no gap is present between the host and the polar disk) by the relative brightness of their structural components: the polar bulge is significantly fainter than the host structure, is often reliably detectable only in deep imaging (e.g., DESI Legacy), and exhibits a relatively shallow central light profile. We note that some apparent PBs may instead be unresolved small PRs (this is discussed in detail in Sect.~\ref{sec:reliability}). True PBs are thought to originate from minor mergers or tidal accretion events \citep{2025A&A...698L..21B}.

Polar halos (PHs) are smooth, symmetrical, oval-shaped outer structures in the form of a halo or shell, extended in the polar direction relative to the host galaxy major axis \citep{2020MNRAS.494.1751M,2022RAA....22k5003M}. They can originate from a collection of unresolved stellar streams, plumes, or shells produced by multiple minor mergers \citep{2005ApJ...635..931B,2010MNRAS.406..744C,2016ApJ...830...62M,2025ApJ...989..107W}. Alternatively, PHs may represent unresolved PRs, or isophotal twists caused by complex, unresolved structures within the host halo \citep{2001ApJ...548...33B}.

Polar dust lanes (PDLs) are distinct dust lanes not aligned with the major axis of the galaxy (often minor‐axis dust lanes), without any prominent stellar counterpart \citep{1978ApJ...226L.115B}. Usually, PDLs are observed through absorption against the host galaxy (typically an elliptical galaxy, \citealt{1987IAUS..127..135B}) and likely arise from prior wet major \citep{1997ApJ...486L..87B} or gas-rich minor \citep{2012MNRAS.423...59S,2015MNRAS.449.3503D} mergers as well as external gas accretion \citep{2005A&A...435...43V}. In addition, PDLs can indicate gaseous polar structures that lack a prominent stellar counterpart, or where dust strongly obscures the stellar ring.

Polar tidal structures (PTSs) are tidal features that visually cross the main body of the host galaxy at a large inclination relative to the host’s major axis. In most cases, the host is a highly inclined system, and the polar structure is double-sided, extending from above to below the galaxy plane. This category includes rings or extended arcs in the polar or strongly tilted direction that exhibit at least one of the following characteristics: a nearly (but not entirely) closed loop; multiple rings or loops; significant asymmetry; an off-centered position relative to the host galaxy; or asymmetrical streaks (in the case of an edge-on orientation) crossing the galaxy body. A subset of these systems displays multiple, non-coplanar loops that wind around the host in a helical fashion, forming a cocoon-like envelope. Most PTSs likely arise from minor mergers or tidal accretion from nearby donor galaxies and some are accompanied by tidally disrupted satellites \citep{2010AJ....140..962M,2024MNRAS.530.4422K,2025A&A...700A.176M}. Some of these systems may represent early stages in the formation of PRs, PBs, or PHs.

Galaxies with extremely tilted or warped stellar disks (denoted as EW) and X-shaped bulges elongated in the vertical direction are sometimes included in studies of polar or tilted structures. Since these galaxies are not the primary focus of this study, we did not specifically target them during the selection process and only included them in cases where they might be confused with real PSGs. Consequently, our catalog is significantly incomplete with respect to these types of objects.

\begin{figure}
    \centering

    \begin{subfigure}[b]{0.48\columnwidth}
        \includegraphics[width=\linewidth]{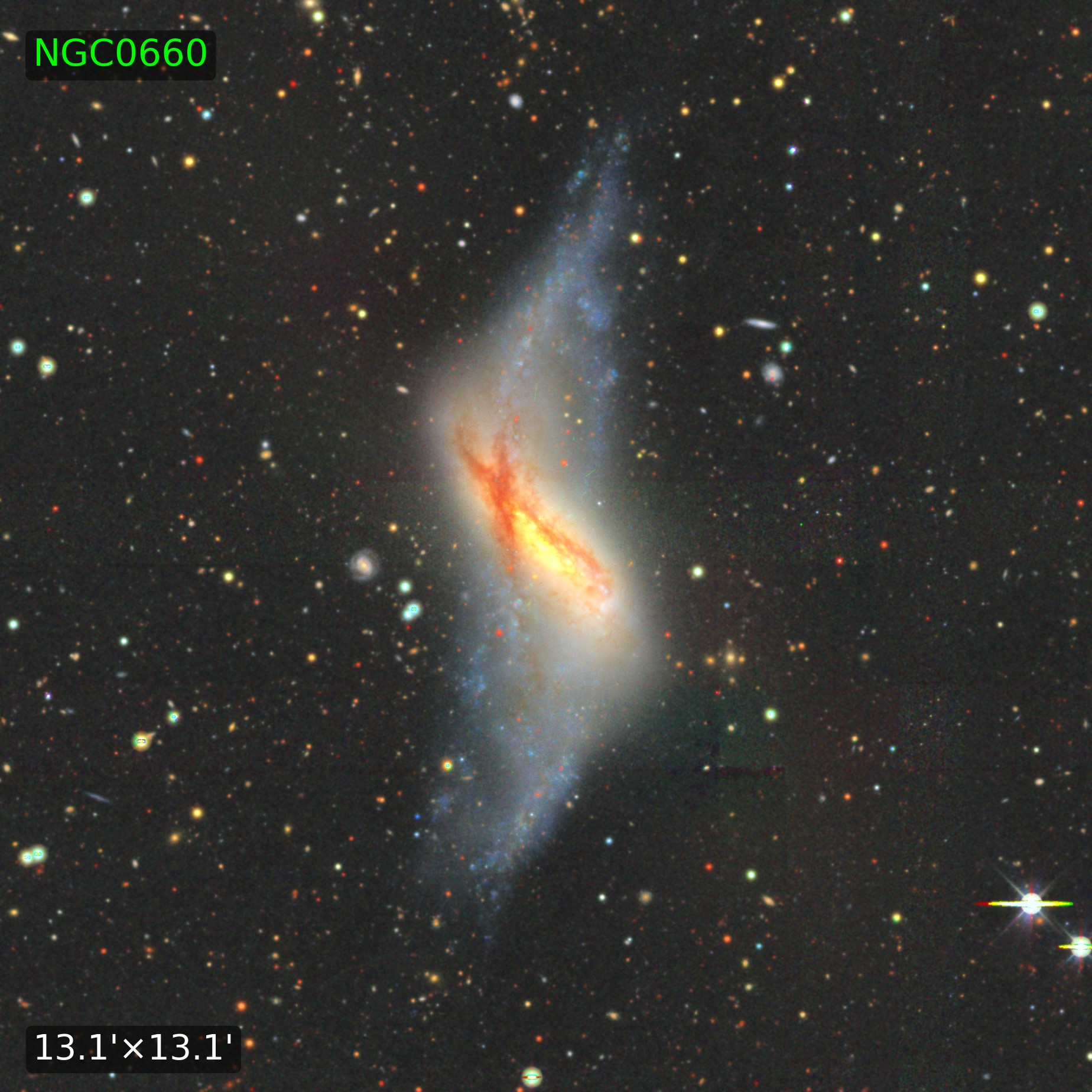}
    \end{subfigure}
    \hfill
    \begin{subfigure}[b]{0.48\columnwidth}
        \includegraphics[width=\linewidth]{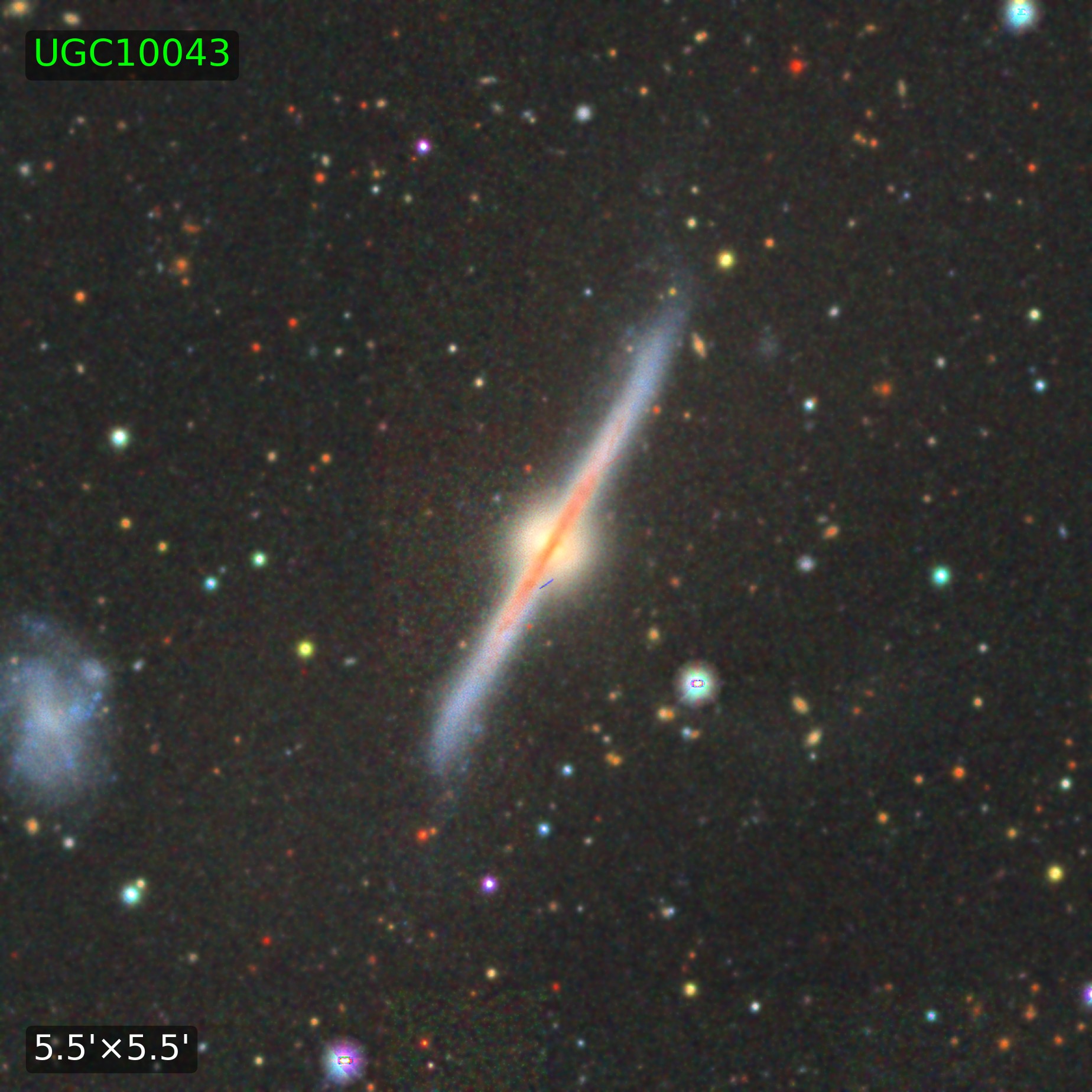}
    \end{subfigure}

    \vspace{1em}

    \begin{subfigure}[b]{0.48\columnwidth}
        \includegraphics[width=\linewidth]{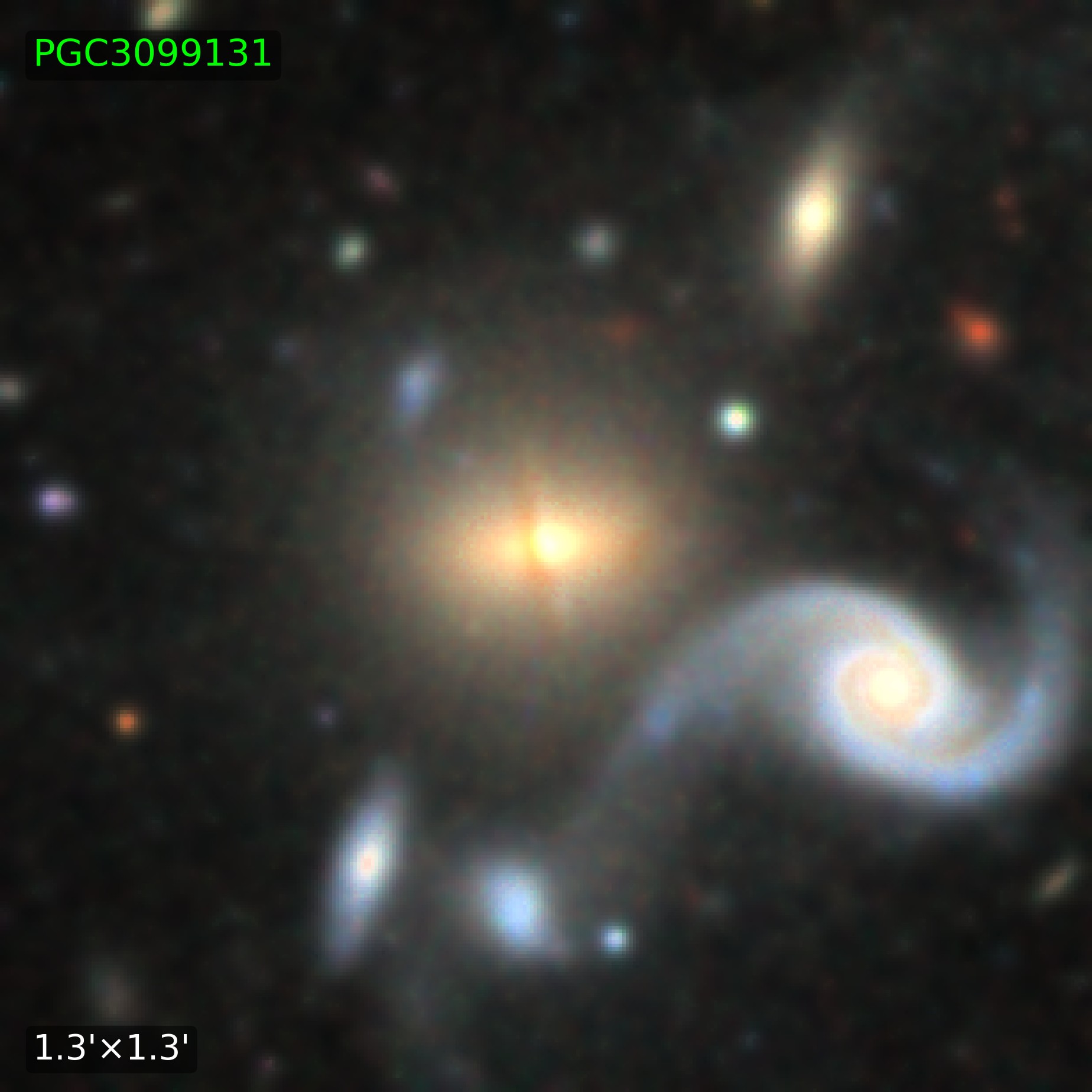}
    \end{subfigure}
    \hfill
    \begin{subfigure}[b]{0.48\columnwidth}
        \includegraphics[width=\linewidth]{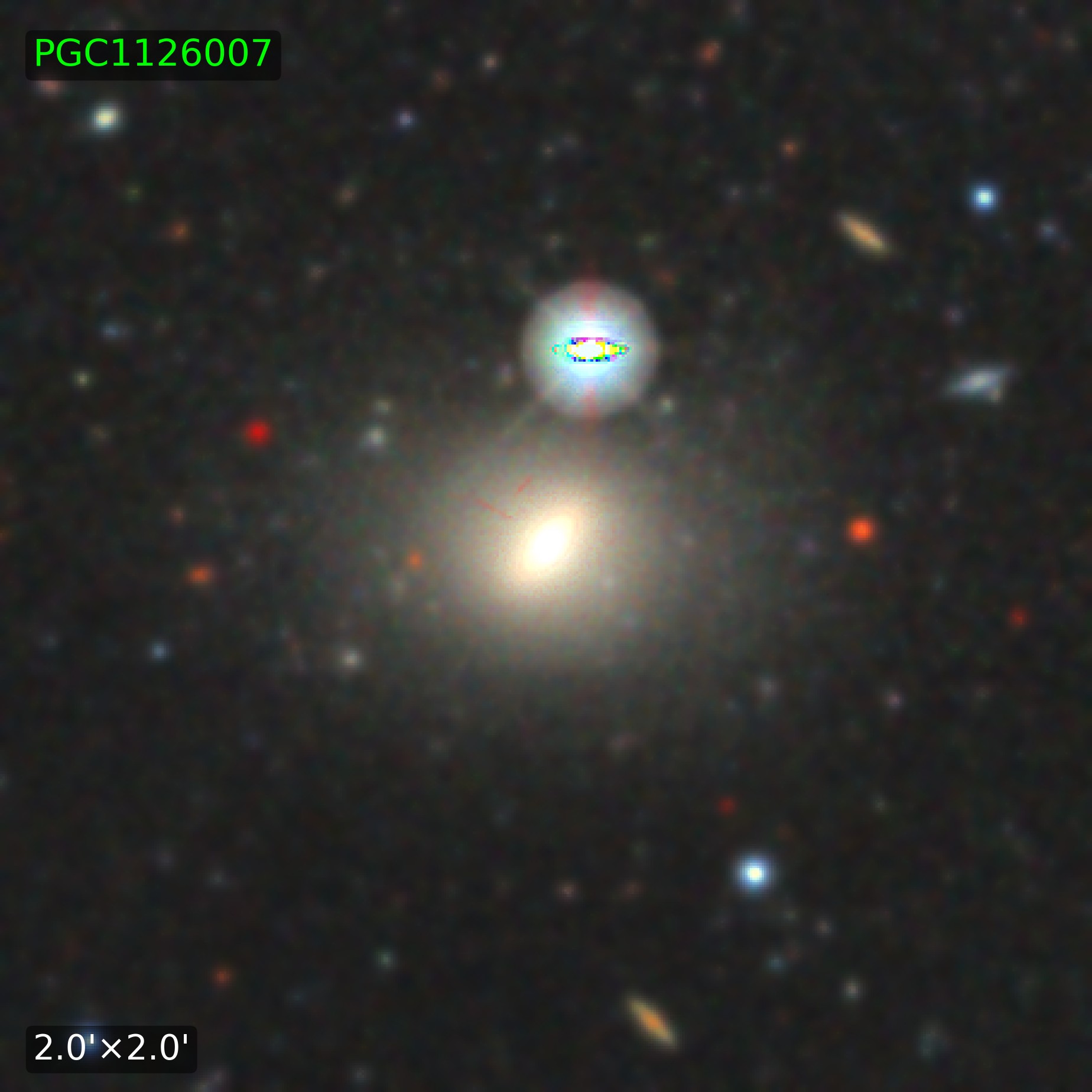}
    \end{subfigure}

    \vspace{1em}

    \begin{subfigure}[b]{0.48\columnwidth}
        \includegraphics[width=\linewidth]{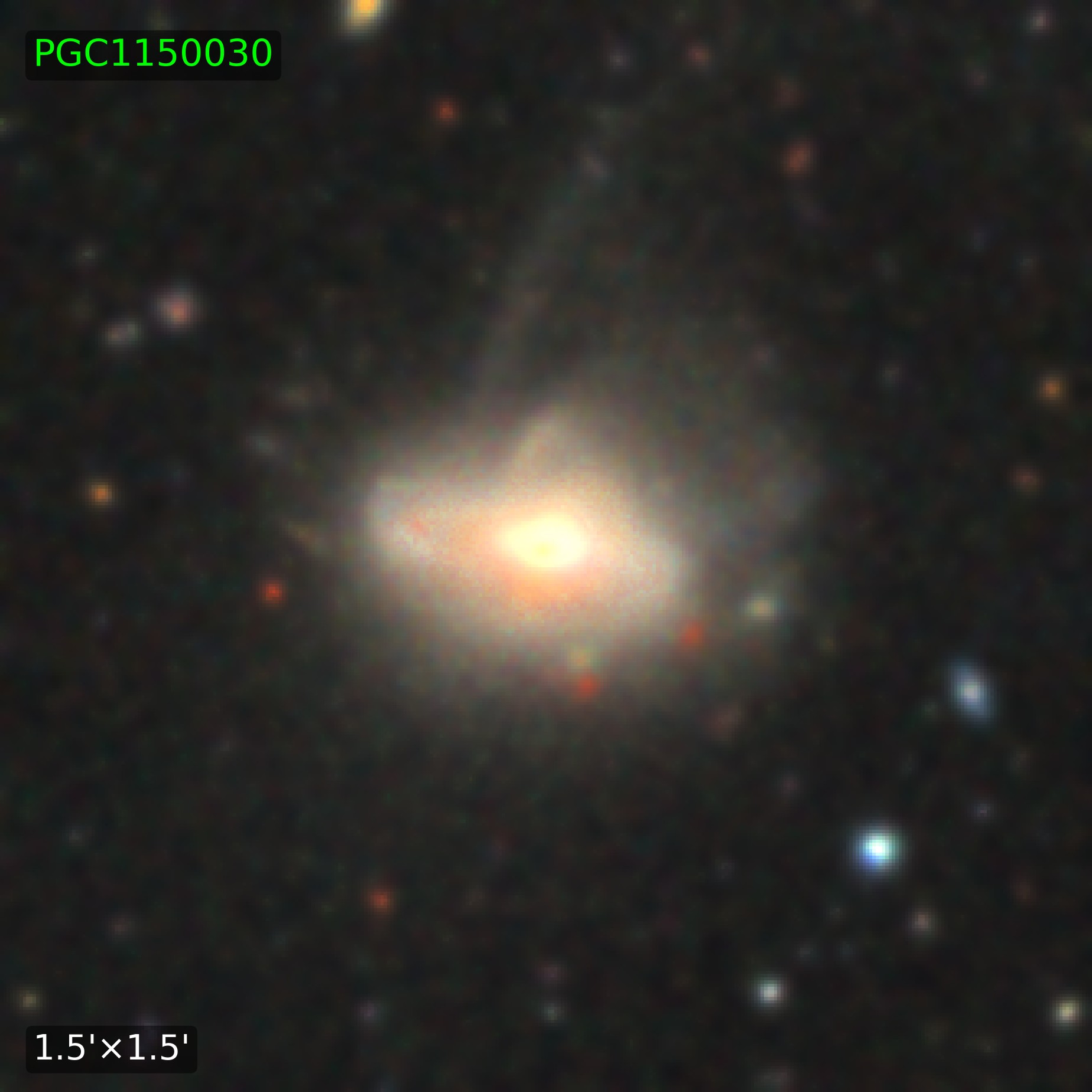}
    \end{subfigure}
    \hfill
    \begin{subfigure}[b]{0.48\columnwidth}
        \includegraphics[width=\linewidth]{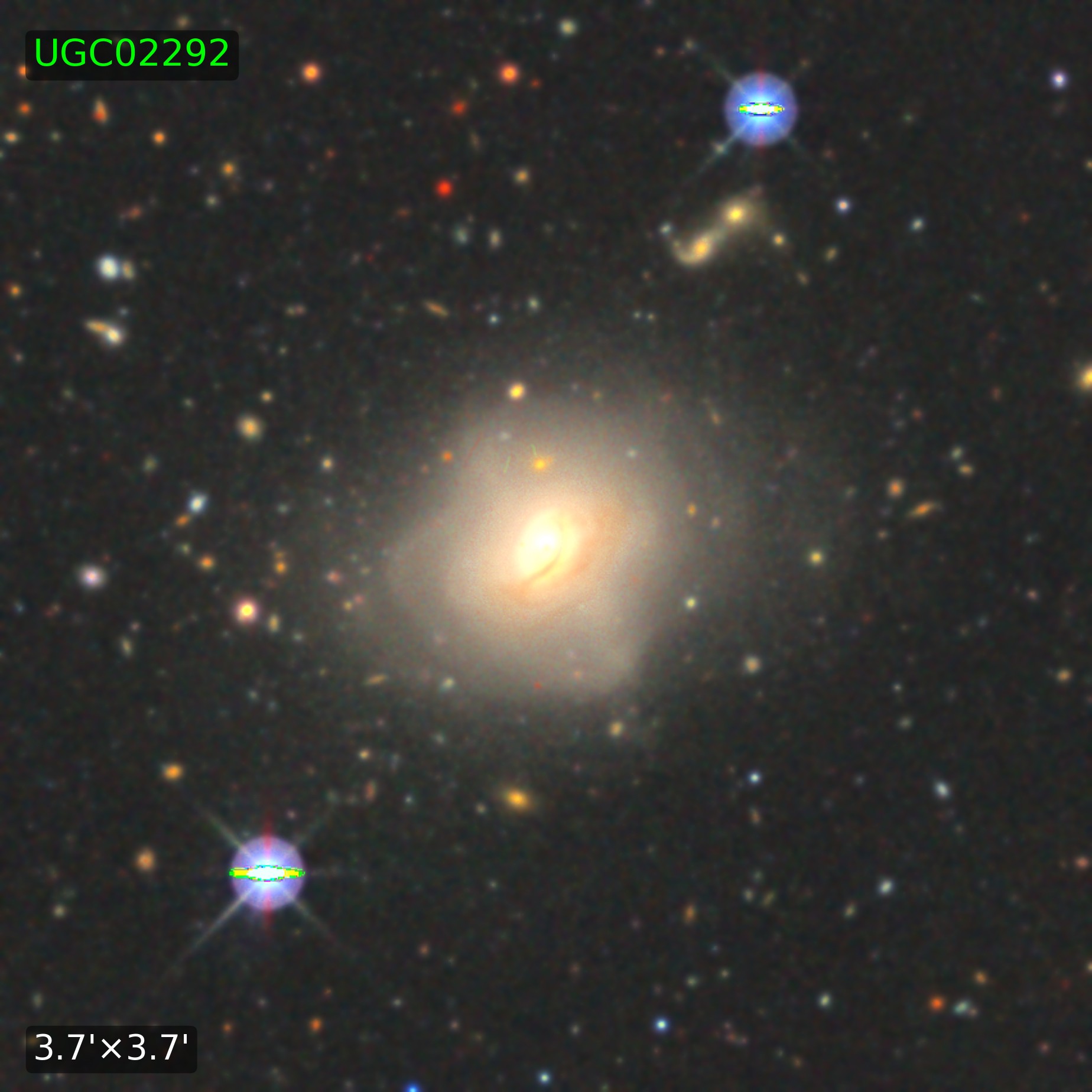}
    \end{subfigure}

    \caption{Examples of PSG subtypes from the literature \citep{1990AJ....100.1489W, 2011MNRAS.418..244M, 2015AstL...41..748R, 2024A&A...681L..15M}. Clockwise from top-left: PR, PB, PH, PTS, PTS, and PDL.}
    \label{fig:subtype_ex}
\end{figure}

Examples of typical galaxies corresponding to each subtype are shown in Fig.~\ref{fig:subtype_ex}. In the selection procedure described below, we first identified PSGs as a general class within the DESI Legacy data, and subsequently classified the selected candidates into PSG subtypes based on the morphological attributes outlined above. It should be noted, however, that this classification is inherently subjective. A quantitative photometric and spectroscopic analysis is required to validate the classification and will be conducted in subsequent papers.

\subsection{Selection from the literature}\label{subsec:lit_selec}

To construct our catalog of PSGs, we first examined previously reported PSGs in the literature through an extensive search of the NASA ADS database\footnote{\url{https://ui.adsabs.harvard.edu/}} for all relevant publications. Our search yielded four large catalogs of PRGs \citep{1990AJ....100.1489W, 2011MNRAS.418..244M, 2019MNRAS.483.1470R, 2024A&A...681L..15M}, described below, as well as many smaller-scale studies of both PRGs and other PSG subtypes.

The first major catalog of PSGs, named the Polar Ring Catalog (PRC), was compiled by \citet{1990AJ....100.1489W}, who examined existing galaxy catalogs for relevant search terms and assembled a list of potentially related galaxies. They then inspected survey plates from the Guide Star Selection System to visually identify PRG candidates and kinematically confirmed as many PRG candidates as possible. The resulting catalog was divided into four categories: ``kinematically confirmed'' (PRC--A, 17 objects), ``good candidates'' (PRC--B, 73 objects), ``possible candidates'' (PRC--C, 67 objects), and ``possibly related systems'' (PRC--D, 23 objects). In total, they classified 157 galaxies as confirmed, good, or possible candidates, and an additional 23 as related structures, yielding 180 potential PSG candidates.  

Upon visual inspection using the higher-fidelity imaging from the DESI Legacy~DR10, we were able to rule out most of the original candidates as false positives, leaving a total of 44 good candidates from the PRC to include in our catalog. Many of the rejected objects in \citet{1990AJ....100.1489W} had already been classified as related structures. For example, NGC\,520 (PRC\,D--44) is an active merger without distinct polar elements, and PGC\,3817123 (PRC\,D--26) shows only a slight warping in one of its spiral arms rather than a complete polar structure. Conversely, NGC\,3718 (PRC\,D--18) clearly exhibits a faint polar dust feature, and we included it in our catalog despite it not being classified as a good or possible PRG candidate by \citet{1990AJ....100.1489W}.

The next large PSG catalog, the SDSS-based Polar Ring Catalogue (SPRC), was compiled by \citet{2011MNRAS.418..244M} using data from the SDSS DR7 \citep{2009ApJS..182..543A} and galaxy identifications from the Galaxy Zoo Project \citep{2011MNRAS.410..166L}. They first examined the Galaxy Zoo forum for reports of unusual or ring galaxies, then compared their preliminary list with the morphological classifications from the Galaxy Zoo to establish criteria for narrowing down additional potential candidates. Using the Galaxy Zoo classification, they visually inspected 41,958 galaxies and identified 275 candidates for PSGs. This sample was divided into 70 best candidates, 115 good candidates, 37 potential face-on PRGs, and 53 related objects.  

After reexamining low-resolution features with the increased photometric depth of DESI Legacy~DR10 (28.5--29\,mag\,arcsec$^{-2}$ versus 26.5\,mag\,arcsec$^{-2}$ in SDSS DR7), we reduced the sample of \citet{2011MNRAS.418..244M} to a total of 150 PSGs. For example, SPRC\,53, which was initially classified as a PRG, is shown by the DESI Legacy~DR10 data to be a highly inclined galaxy accompanied by two symmetrically located objects above and below its disk. Likewise, SPRC\,44 appears as an edge-on galaxy with two galaxies projected above and below its bulge.

The third significant sample of PSGs  was compiled by \citet{2019MNRAS.483.1470R}, who further examined galaxies classified within the Galaxy Zoo Project in an effort to increase the known number of PRGs. They expanded the sample of \citet{2011MNRAS.418..244M} by 31 galaxies. After removing duplicates already included in previous catalogs and reexamining the sample with improved visual clarity and greater photometric depth from the DESI Legacy~DR10, we reduced that number to 25 good PSG candidates. We excluded galaxies such as PGC\,907838, which is a normal barred spiral galaxy.

The most recent PSG catalog was compiled by \citet{2024A&A...681L..15M}, who searched for PSGs in data from SDSS Stripe\,82 \citep{2016MNRAS.456.1359F}, a region of the sky observed with significantly greater photometric depth than SDSS \citep[up to 1.8\,mag deeper,][]{2019MNRAS.486..390B}. These data were further supplemented by examinations of galaxies in the DESI Legacy DR9 \citep{2019AJ....157..168D} and the Hyper Suprime-Cam Subaru Strategic Program (HSC-SSP) DR3 \citep{2022PASJ...74..247A}. As noted above, this catalog was also the first to include multiple types of polar structures as subtypes within a single broader structural family. \citet{2024A&A...681L..15M} identified a total of 53 good candidates for PRGs, 49 good candidates for other types of PSGs, and about 50 related objects. After further examination and restricting our analysis to galaxies contained in the DESI Legacy~DR10, we reduced that number to a total of 103 strong PSG candidates. 

Beyond the catalogs described above, we also examined other studies and added additional good candidates to our catalog. After identifying papers focused on an analysis of PRs and other types of polar structures, we visually inspected the galaxies mentioned therein, using the DESI Legacy~DR10 to determine their inclusion in our catalog. The studies we considered included individual identifications such as the PRG DES\,J024008.08$-$551047.5 \citep{2024A&A...681A..35A}, as well as larger samples such as \citet{2003A&A...408..873C}, featuring a sample of galaxies with inner polar disks, and \citet{2015AstL...41..748R}, featuring a sample of galaxies with PBs. In total, we identified 95 potential PSG candidates, which were then visually examined in the DESI Legacy~DR10. We ultimately excluded galaxies without clearly identifiable external polar structures \citep[studied, for example, in][]{2003A&A...408..873C,2016AJ....152...73S}, as well as those exhibiting only gaseous polar structures \citep[e.g.,][]{2023MNRAS.525.4663D}, to maintain consistency with our general SGA selection criteria. Overall, we added 20 additional good candidates to our catalog.

By limiting our selection to PSGs present in the DESI Legacy~DR10 and eliminating duplicates across studies, we identified 727 PSG candidates from the literature. After visually inspecting these objects in the DESI Legacy~DR10, we reduced the sample to 342 good candidates. This reduction reflects both the quality of imaging in DESI Legacy and our decision to restrict the catalog to visually confirmed external polar structures. As part of our visual examination, we reclassified each PSG following the definitions in Sect.~\ref{subsec:nom} (noting that in most cases, our classification matches the original). Table~\ref{tab:litpsgs} summarizes the relationship between earlier catalogs and our compilation.

\begin{table}
    \caption{Number of potential PSG candidates identified in the literature (N$\mathrm{tot}$) compared to the number of good candidates in the DESI Legacy~DR10 that were added to our catalog (N$\mathrm{good}$).}
    \begin{tabularx}{0.5\textwidth} { 
   >{\raggedright\arraybackslash}X 
   >{\raggedright\arraybackslash}p{0.7cm} 
   >{\raggedright\arraybackslash}p{0.7cm} 
   >{\raggedright\arraybackslash}X  }
        \hline
        Catalog & N$_{tot}$ & N$_{good}$\tablefootmark{a} & Data Source\tablefootmark{b}\\
        \hline
        \hline
        PRC \citep{1990AJ....100.1489W} & 180 & 44 & Uppsala General Catalog of Galaxies, ESO Catalog of Galaxies, Southern Atlas of Peculiar Galaxies\\
        \hline
        SPRC \citep{2011MNRAS.418..244M} & 275 & 150 & SDSS DR7\\
        \hline
        \cite{2019MNRAS.483.1470R} & 31 & 25 & SDSS DR7\\
        \hline
        Mosenkov et al. (2024a) & 152 & 103 & SDSS Stripe 82, DESI Legacy DR9, HSC-SSP DR3\\
        \hline
        Other & 95 & 20 & Various\\
        \hline
        \hline
        Total & 727 & 342 & \\
        \hline
        \end{tabularx}
        \tablefoottext{a}{N$_\mathrm{good}$ accounts for the removal of duplicates and galaxies not present in DESI Legacy.}
        \tablefoottext{b}{Data Source refers to the survey or catalog originally used to compile each sample.}
    \label{tab:litpsgs}
\end{table}

\subsection{Using convolutional neural networks}\label{subsec:cnn}

The application of deep learning to astronomical image analysis has developed rapidly in recent years, with CNNs demonstrating exceptional performance in the morphological classification of galaxies and other celestial objects \citep{2019Ap&SS.364...55Z,2021MNRAS.506..659C,2022MNRAS.513.1581W,2024A&A...683A..42C}. The successful application of CNNs to tasks such as galaxy morphology classification \citep{2019MNRAS.484...93D,2022MNRAS.509.3966W}, star–galaxy separation \citep{2017MNRAS.464.4463K,2024RAA....24i5012Z}, and transient detection \citep{2019MNRAS.489.3582D,2025ApJ...987..105K}, demonstrates their value for large-scale survey data, where a manual classification would be impractical.

Recent works have explored the use of deep learning for identifying PRGs in large imaging surveys. Although these efforts are still in the early stages, they have revealed both the potential and the current limitations of automated approaches to PRG identification. For instance, \citet{2025A&A...702A.258D} developed the first deep-learning method for identifying PRGs using SDSS imaging data. Starting from existing PRG catalogs, they constructed a small training set of 87 good PRGs, augmented through image segmentation, ensemble learning, and, most effectively, transfer learning with synthetic images generated via \texttt{GALFIT}. Although this approach has demonstrated the potential of deep learning for recognizing PR-like morphologies, it ultimately produced only three convincing PRG detections in the SDSS at $z < 0.1$. \citet{2025RASTI...4af043K} presented a sophisticated neural-network architecture optimized for identifying PRGs, showing that more specialized models and larger, well-curated training datasets will be essential for successful large-scale PSG detection in future surveys. However, their CNN has not yet been applied to discover new PRGs. This work takes the application of deep learning for galaxy classification one step further by training a CNN to differentiate PSGs (of all types) from typical galaxies. Although PSG morphologies are diverse, they are distinct as a class compared to galaxies without polar structures, with the major orthogonally oriented structure standing out as the defining feature.

\begin{figure}
    \centering
    \includegraphics[width=0.7\linewidth]{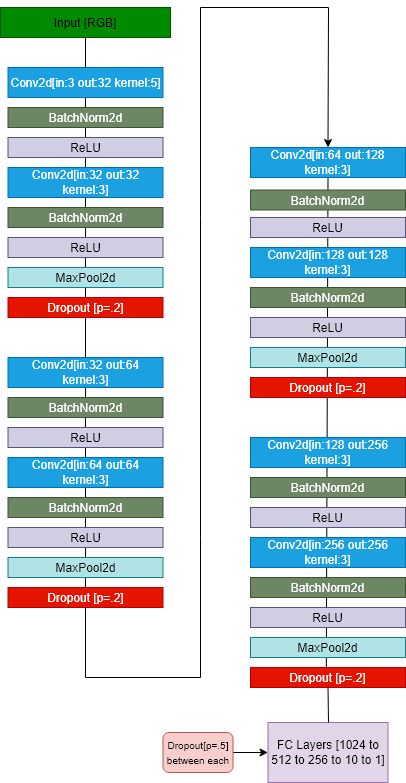}
    \caption{Architecture of the prototype custom CNN trainer (CNN~1).}
    \label{fig:cnnArch}
\end{figure}

To assist in the identification of PSGs, we developed two CNN architectures trained to classify galaxy images. The first CNN (CNN~1) employed a custom binary classification model designed to distinguish PSGs from non-PSGs. The model consisted of four convolutional blocks, each containing two Conv2D layers with batch normalization, ReLU activation, max-pooling, and dropout layers to reduce overfitting. The architecture of CNN~1 is shown in Fig.~\ref{fig:cnnArch}. We significantly expanded the training sample compared to \citet{2025A&A...702A.258D} and \citet{2025RASTI...4af043K} to account for the increased diversity of objects introduced when including all PSG subtypes. The network was trained using 11,400 augmented images from DESI Legacy~DR10, including 588 PSGs and 2,262 non-PSGs, with each image resized to $256\times256$\,px and converted into a tensor representation. The training utilized a BCE loss function with the Adam optimizer and a dynamic learning-rate scheduler, initially set to $5\times10^{-4}$ and reduced by 30\% every 10 epochs. The training proceeded for 200~epochs.

To counter the strong class imbalance between PSGs and non-PSGs, an image augmentation (horizontal and vertical flips) was applied to expand the PSG dataset fourfold. The validation used 84 PSG and 238 non-PSG images, yielding an overall validation accuracy of 93.4\%. When applied to the full SGA sample of 383,620 galaxies, the trained CNN~1 identified 737 PSG candidates, corresponding to a 99.8\% reduction of the total sample size. The manual inspection confirmed 21 of these as convincing PSG candidates, with 7 PTSs, 4 PBs, 4 PDLs, 3 PRs, and 3 PHs. Although the model produced many false positives, it demonstrated the feasibility of CNN-based preselection in filtering vast survey data for rare morphological types.

\begin{figure}
    \centering
    \includegraphics[width=1\linewidth]{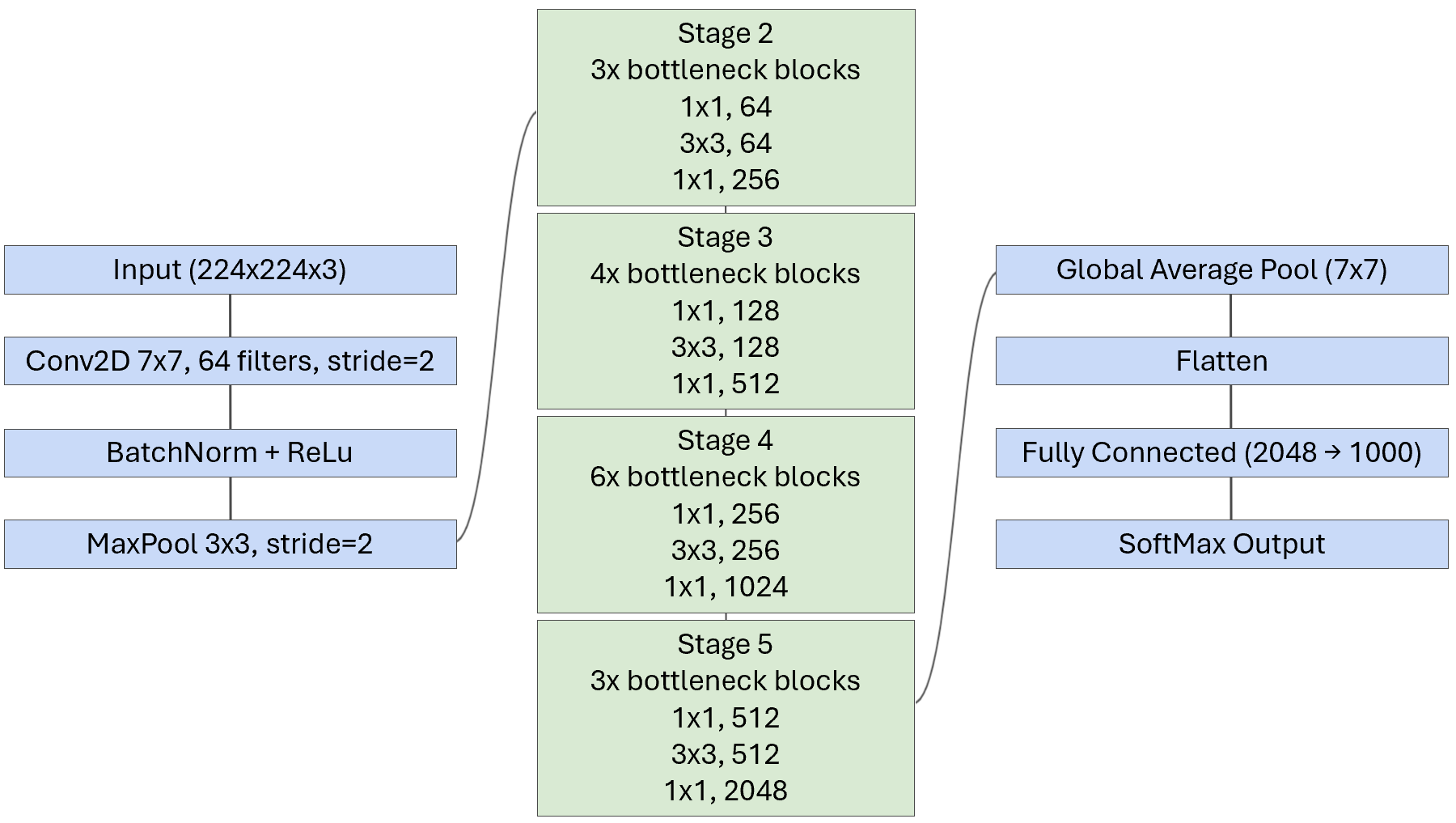}
    \caption{Architecture of the updated CNN for PSG selection (CNN~2). }
    \label{fig:new_cnn_arch}
\end{figure}

The second version of the CNN (CNN~2)  employed a more advanced architecture based on ResNet50 with transfer learning to enable multi-class classification. Instead of a binary PSG or non-PSG output, the model classified galaxies into five morphological types: four regular types of galaxies without peculiarities (spiral, elliptical, lenticular, and irregular, all based on our own morphological classification of $\sim20,000$ randomly selected galaxies from the SGA) and PSGs. Major mergers were excluded from consideration. The use of a pretrained ResNet50 backbone, combined with transfer learning, allowed the network to leverage low-level feature recognition while retraining higher layers for PSG-specific features. The training was conducted in two phases: first with the convolutional backbone frozen and then with fine-tuning at a reduced learning rate. The training and testing datasets were refined using the results of visual classification since CNN~1. CNN~2 was trained on a sample of 262 PSGs and 7431 non-PSGs, with 38 PSGs and 1066 non-PSGs set aside for testing. Data augmentation was expanded to include random rotations, flips, and noise injection. The architecture of CNN~2 is shown in Fig.~\ref{fig:new_cnn_arch}.

\begin{figure}
    \centering

    \begin{subfigure}[b]{0.48\columnwidth}
        \includegraphics[width=\linewidth]{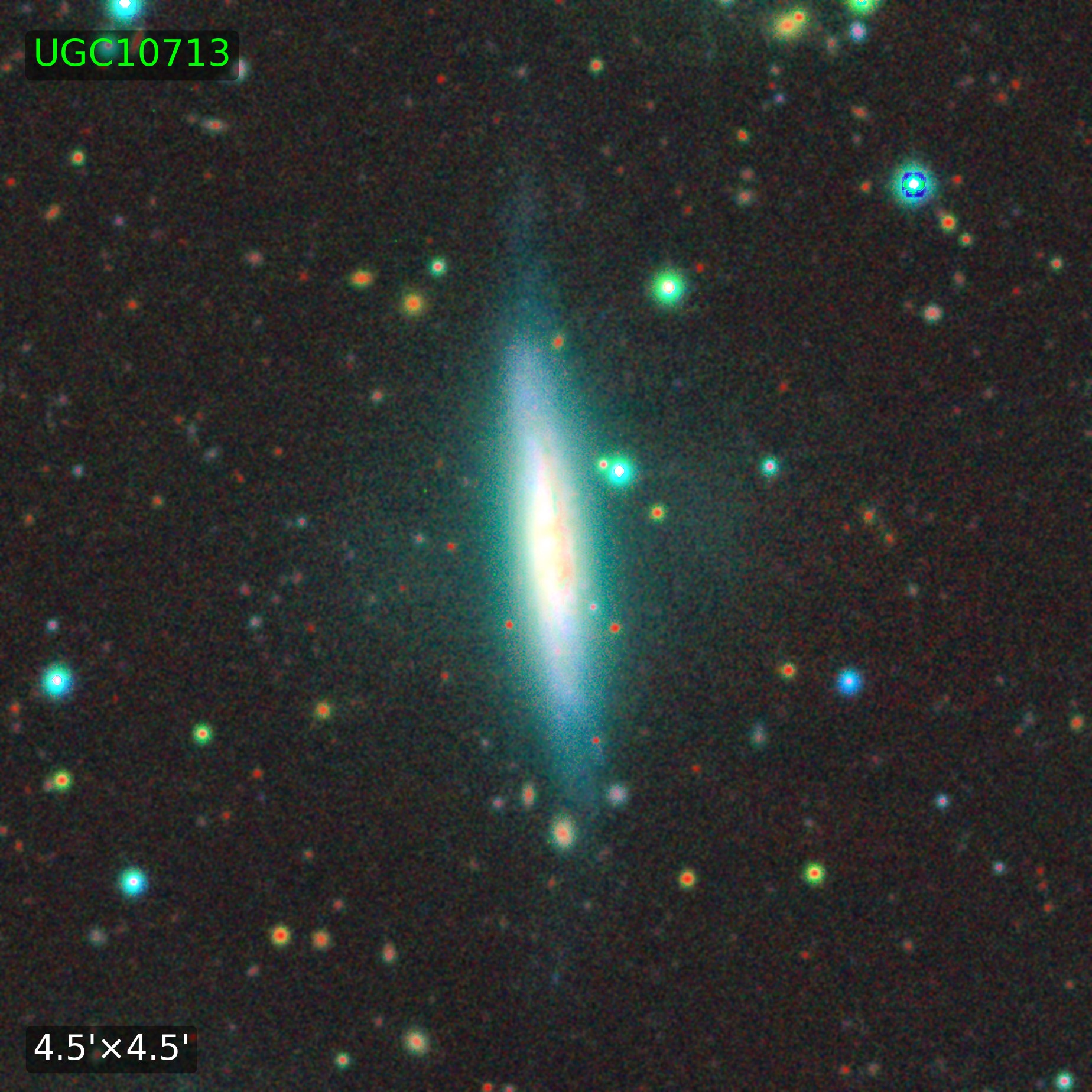}
    \end{subfigure}
    \hfill
    \begin{subfigure}[b]{0.48\columnwidth}
        \includegraphics[width=\linewidth]{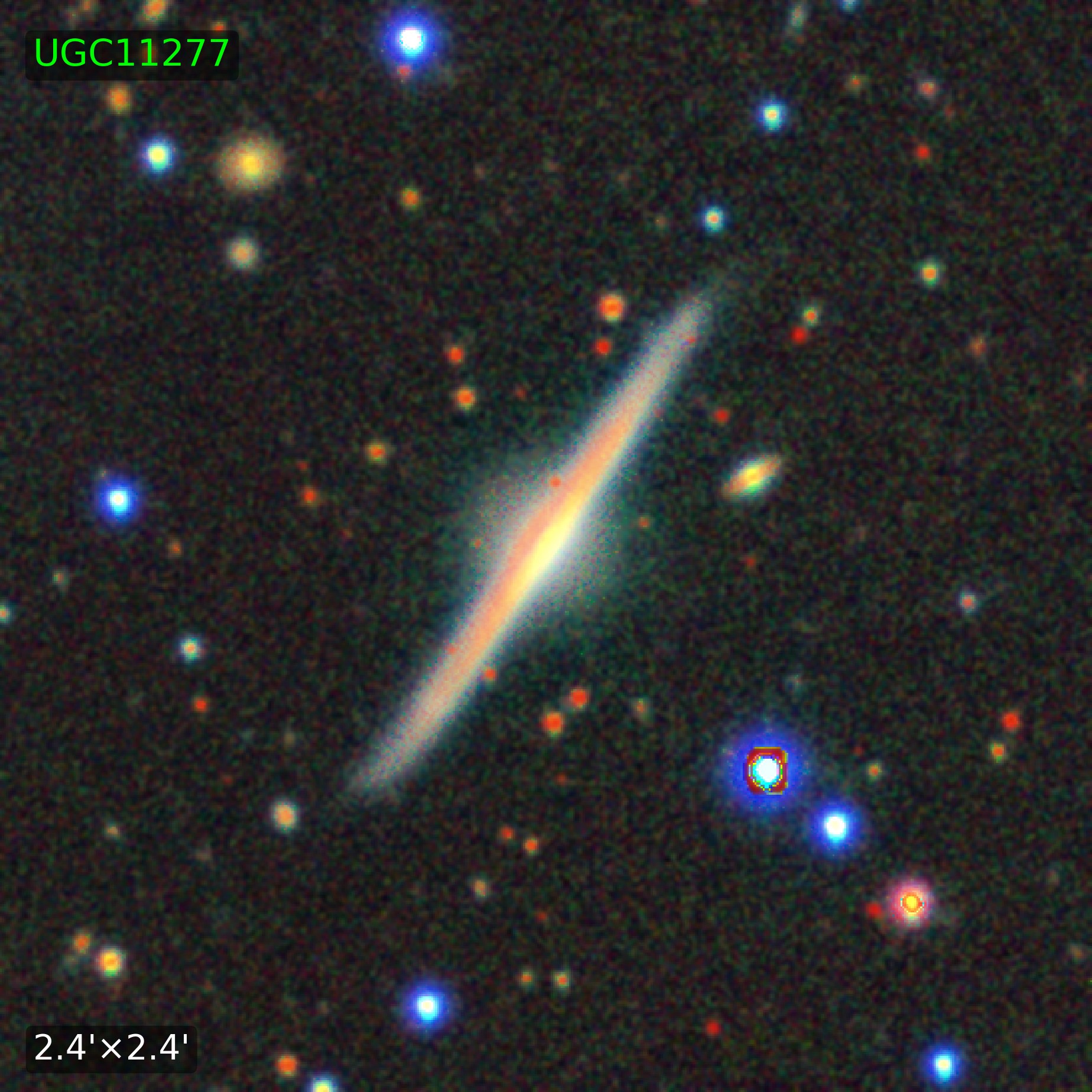}
    \end{subfigure}

    \vspace{1em}

    \begin{subfigure}[b]{0.48\columnwidth}
        \includegraphics[width=\linewidth]{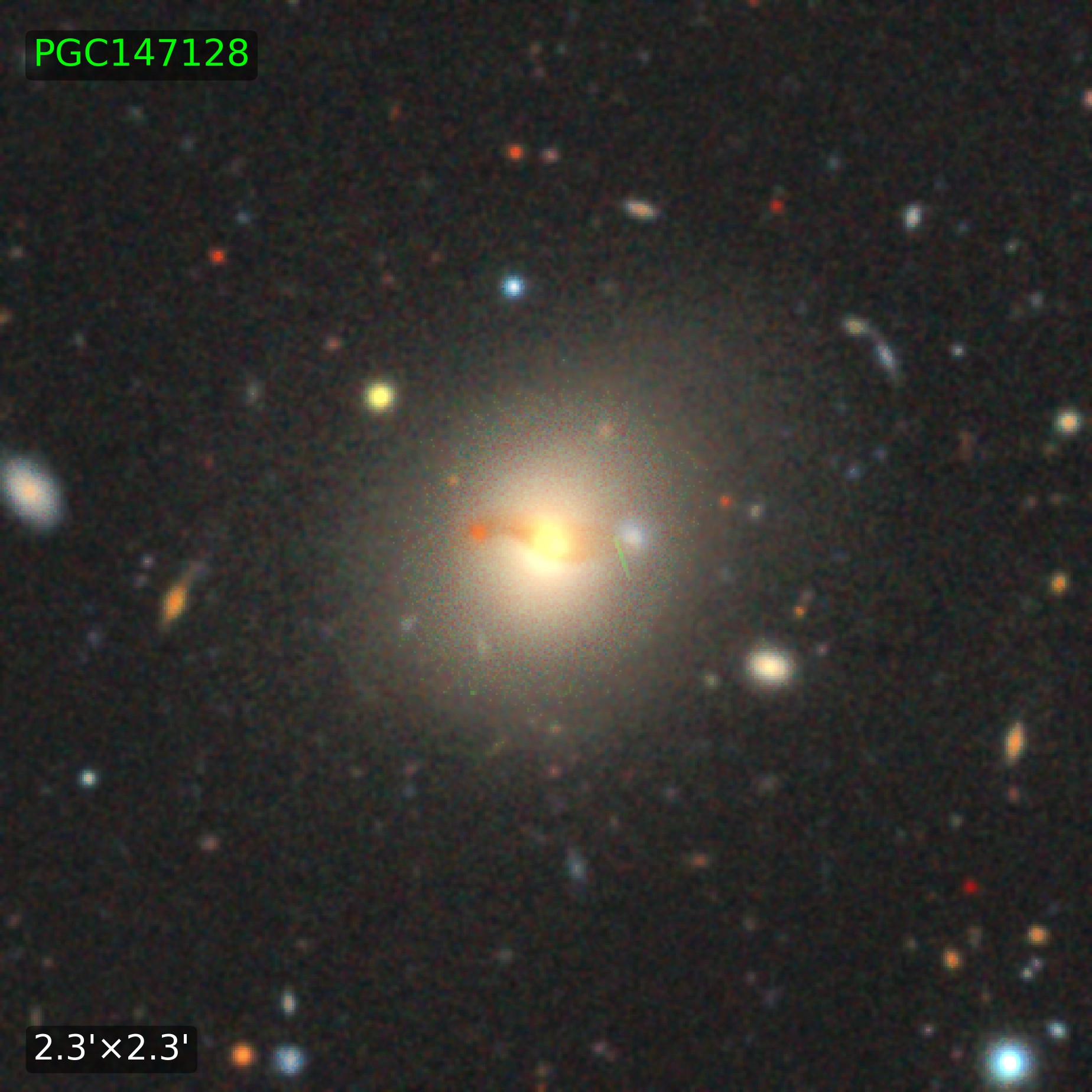}
    \end{subfigure}
    \hfill
    \begin{subfigure}[b]{0.48\columnwidth}
        \includegraphics[width=\linewidth]{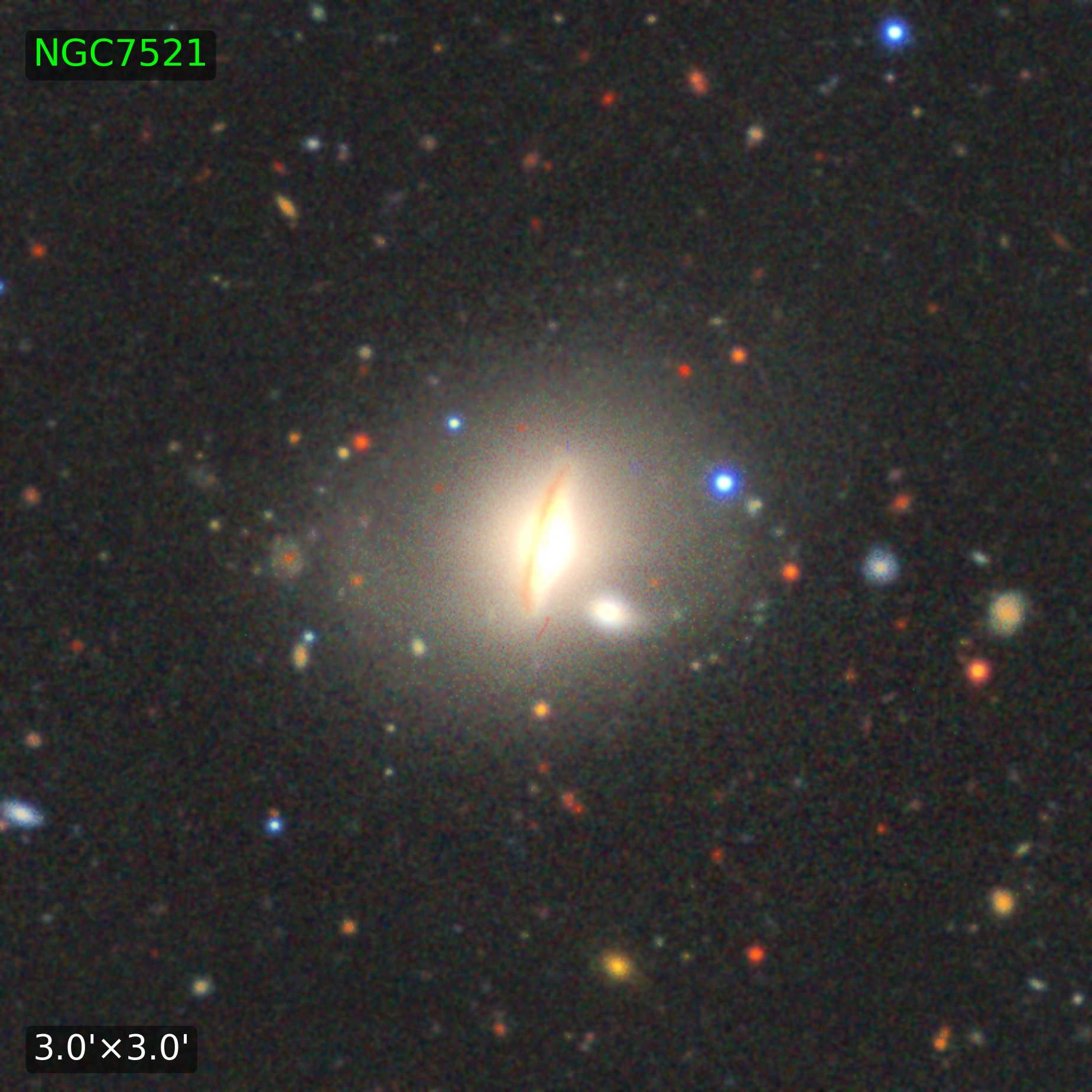}
    \end{subfigure}

    \caption{Examples of PSGs selected by CNN~2. Clockwise from top left: PR, PB, PTS, PDL.}
    \label{fig:cnnpsgs}
\end{figure}

CNN~2 identified 1,505 PSG candidates from the SGA, of which only 43 galaxies appeared to be good candidates for PSGs. Of the 43 good PSG candidates found by CNN~2, 24 were PTSs, 15 were PRs, and 4 were PBs. The confusion matrix is shown in Table~\ref{tab:cnnuncert}. CNN~2 had a recall of 97.4\% and a precision of 92.5\% for the PSG class, indicating that nearly all true PSGs were correctly recovered, with only minor contamination from irregular galaxies. Representative examples of CNN-selected PSGs are shown in Fig.~\ref{fig:cnnpsgs}.

\begin{table}
    \centering
    \caption{Confusion matrix of the CNN~2 training.}
    \begin{tabularx}{0.5\textwidth}{  X  X  X  X  X  X  }
        \hline
        Galaxy type & E & Irr & S0 & PSG & S\\
        \hline
        \hline
        E & 195 & 0 & 15 & 0 & 0\\
        \hline
        Irr & 0 & 68 & 0 & 3 & 0\\
        \hline
        S0 & 49 & 0 & 201 & 0 & 1\\
        \hline
        PSG & 0 & 0 & 1 & 37 & 0\\
        \hline
        S & 2 & 49 & 29 & 0 & 454\\
        \hline
    \end{tabularx}
    \tablefoot{The column headers indicate the true categories: elliptical (E), irregular (Irr), lenticular (S0), PSG, and spiral (S). The rows show the CNN~2 predictions.}
    \label{tab:cnnuncert}
\end{table}

Both CNNs demonstrate that deep-learning  methods can dramatically accelerate the identification of rare morphological galaxy types when combined with visual verification. However, many factors must be considered when searching for PSGs using deep learning. The main challenge arises from the intrinsic rarity of PSGs, which severely limits the size and diversity of available training samples. In addition, their morphologies are highly heterogeneous, ranging from well-defined PRs to faint PHs and barely visible PDLs. This variety makes it difficult for both human and machine classifiers to define consistent feature boundaries. The situation is further complicated by differences in spatial resolution and depth between imaging surveys: as we peer deeper into space to discover more distant objects, polar structures become smaller in angular size and fainter. Projection effects, confusion with stellar bars, rings, or tidal debris, and contamination from overlapping galaxies can also lead to significant misclassifications. 

Overcoming these limitations will require larger, more balanced training sets, realistic synthetic data, and algorithms capable of capturing subtle and multiscale morphological patterns. Future versions of our models will integrate multi-label classification to capture overlapping features between PSG subtypes (e.g., PRs, PBs, and PDLs), as well as active learning strategies to iteratively retrain the network with newly confirmed candidates.

\subsection{Human selection and final classification}\label{subsec:human_selec}

After selecting strong PSG candidates from the literature and identifying promising new candidates using CNNs, we finalized our catalog through an intensive visual inspection of all galaxies from the SGA, as well as from two catalogs of edge-on galaxies: EGIS \citep[][5,747 galaxies]{2014ApJ...787...24B} and EGIPS \citep[][16,551 galaxies]{2022MNRAS.511.3063M}. Edge-on galaxies are ideal systems for identifying PSGs among disk galaxies. Detecting PSGs in such samples provides an important test for estimating their occurrence rate in the Local Universe. In addition, we incorporated the largest catalog of ringed galaxies to date, compiled by \citet{2025ApJS..280...11C}, which includes 62,962 galaxies selected from DESI Legacy using semi-supervised deep learning.

The manual search combined the efforts of a professional astronomer, trained graduate students, and undergraduate research assistants. In addition, two groups of undergraduate students in a second-year astronomy course were specially trained in galaxy classification and assisted in classifying images as part of a course project. Each participant examined at least 10,000 galaxy images for the presence of polar structures. All galaxy images were reviewed at least twice by different classifiers. The selected candidates were subsequently reexamined multiple times using our enhanced RGB images and the DESI Legacy Viewer to minimize the rate of misclassification.

The identification of PSGs presents several challenges that can complicate both visual and automated classification. One of the most common sources of confusion is the misclassification of barred or resonance ringed galaxies, whose in-plane inner or outer rings may mimic polar structures when viewed at certain orientations. Face-on and moderately inclined spiral galaxies are particularly problematic, as their geometry makes it difficult to distinguish genuine polar components from internal rings, tightly wound spiral arms, or disk warps. In more distant systems, polar structures appear less sharply defined due to their smaller angular extent, increased cosmological surface-brightness dimming (proportional to $(1+z)^3$ if surface brightnesses are expressed in AB magnitudes, \citealt{2020ApJ...903...14W}), and the resulting decrease in the signal-to-noise ratio. Low spatial resolution, projection effects, image artifacts, and contamination from overlapping foreground or background galaxies can further obscure faint polar features. In addition, strong dust lanes and disturbed morphologies (often associated with mergers or starburst activity) can mask or distort polar components, leading to ambiguous classifications even in deep imaging data. Throughout our visual classification, we strove to account for all these effects (see Sect.~\ref{sec:discussion}); however, some selection errors are inevitable, particularly given the wide morphological diversity of PSGs and the large sizes of the samples under review.

While the influence of these factors may result in some false positives in our final catalog, the iterative selection process was designed to minimize both false positives and missed PSG candidates. Furthermore, classifiers evaluated the quality of each PSG candidate during the selection process. Initially, all potential PSG candidates were retained, including galaxies with asymmetric, blurry, or extremely faint polar structures. After the first round of selection, the sample was cross-checked against known PSG candidates from the literature to identify newly discovered PSGs. The first sample of potential PSG candidates was then reviewed for quality, with only the strongest candidates retained in the final catalog.

It is important to note that during our visual selection we also identified several objects consisting of two regular, undisturbed galaxies without any associated LSB features, indicating that these systems are likely chance line-of-sight overlaps. Examples of such cases and a discussion of their potential fraction within our PSG catalog are provided in Sect.~\ref{subsec:proj_effects}.

\begin{figure}
    \centering

    \begin{subfigure}[b]{0.48\columnwidth}
        \includegraphics[width=\linewidth]{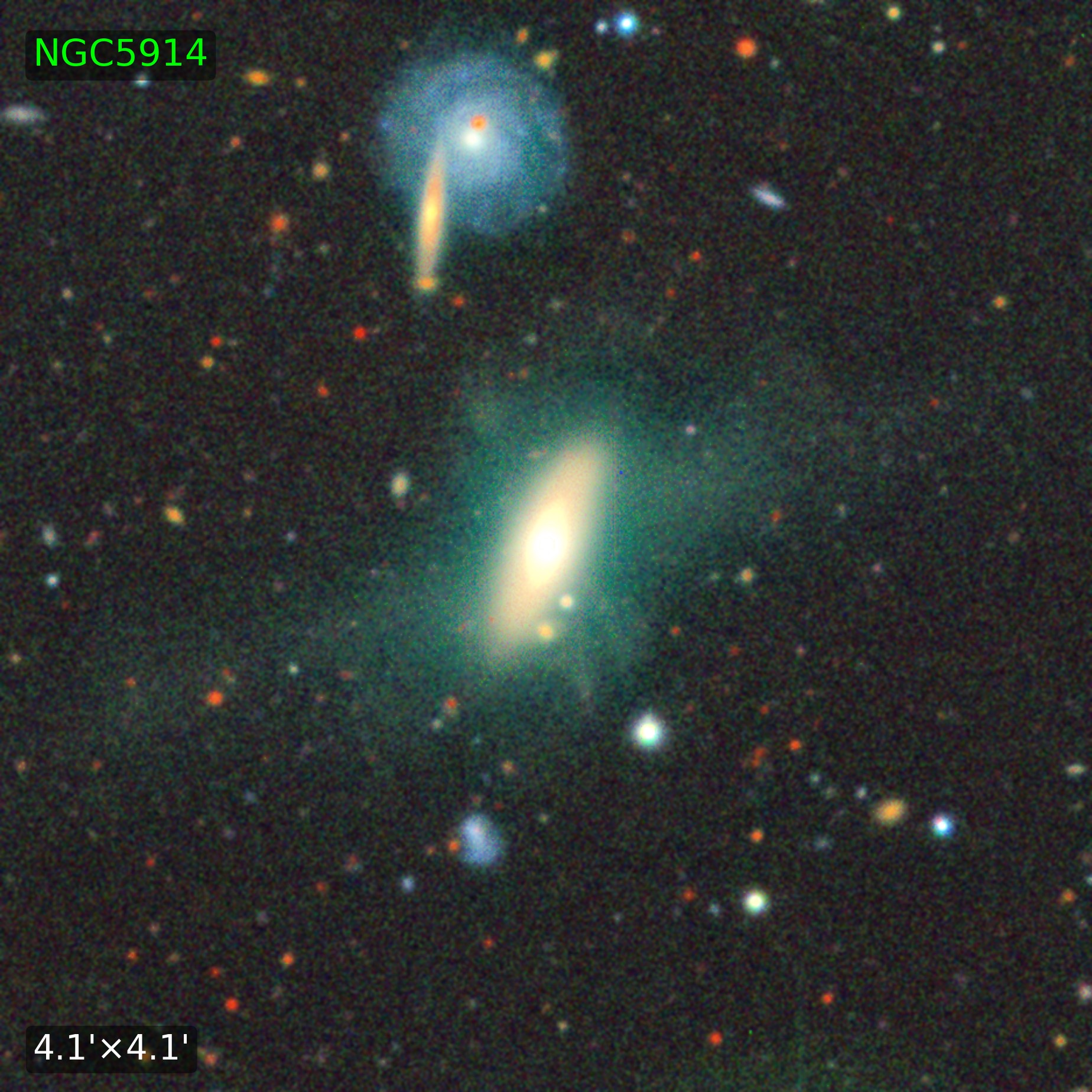}
    \end{subfigure}
    \hfill
    \begin{subfigure}[b]{0.48\columnwidth}
        \includegraphics[width=\linewidth]{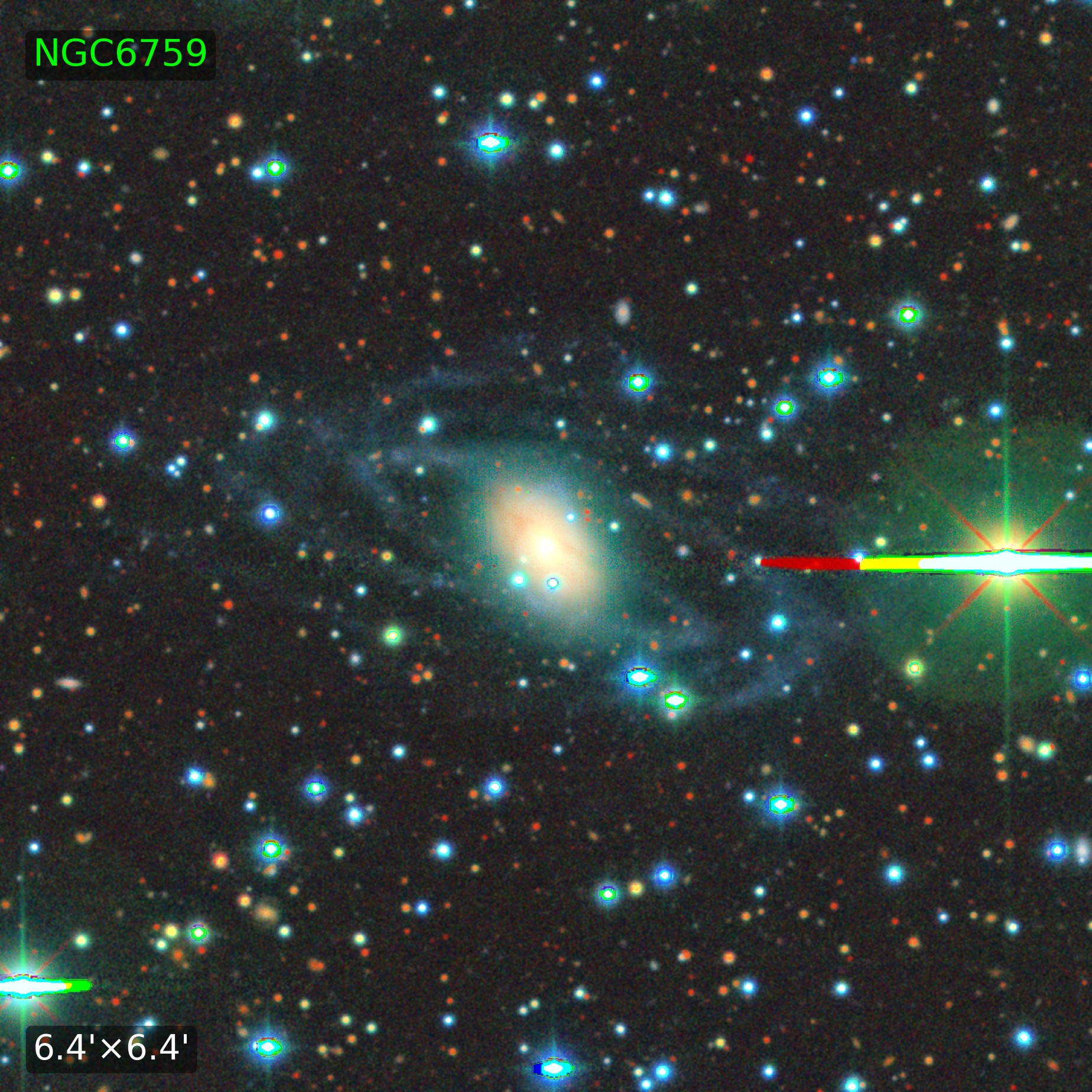}
    \end{subfigure}

    \vspace{1em}

    \begin{subfigure}[b]{0.48\columnwidth}
        \includegraphics[width=\linewidth]{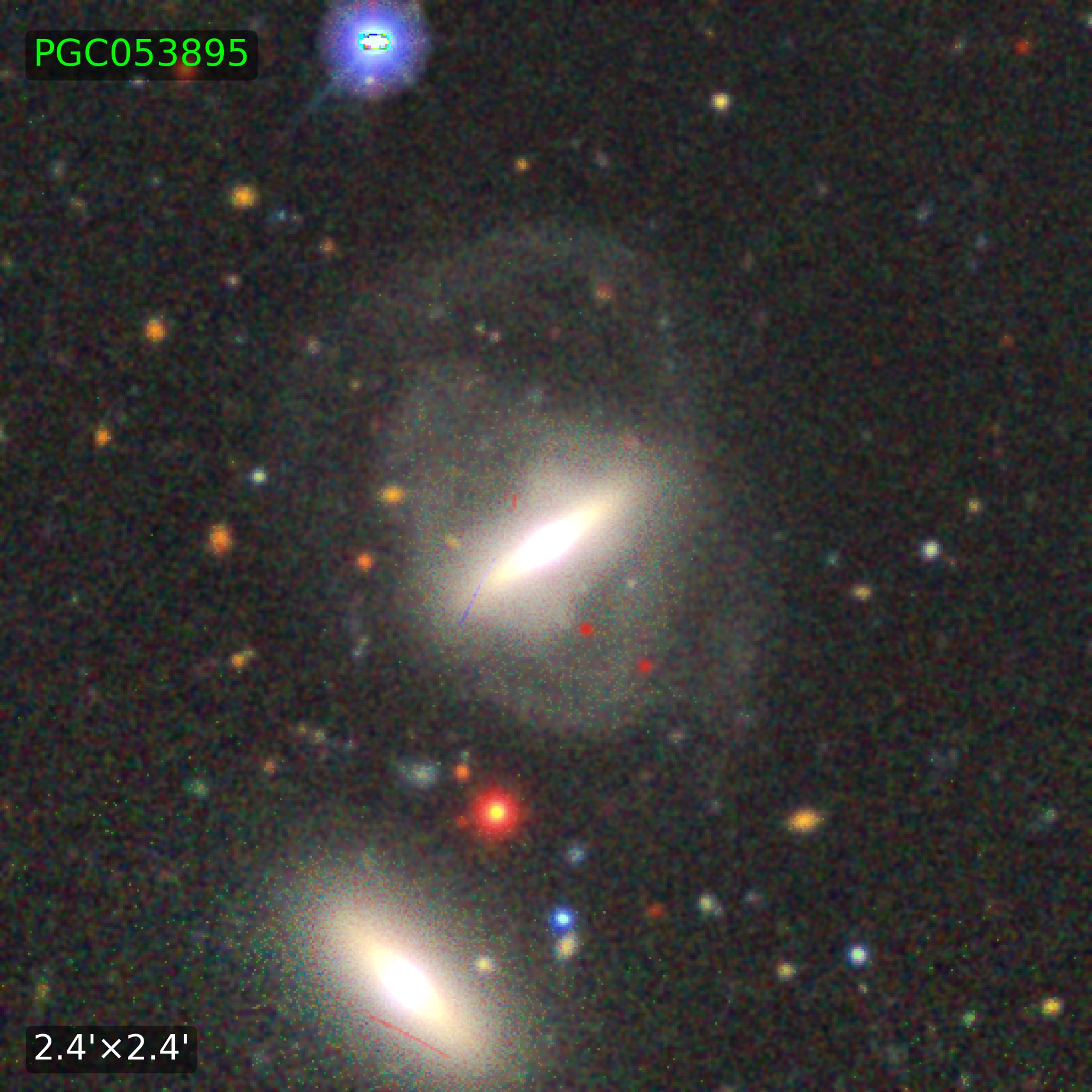}
    \end{subfigure}
    \hfill
    \begin{subfigure}[b]{0.48\columnwidth}
        \includegraphics[width=\linewidth]{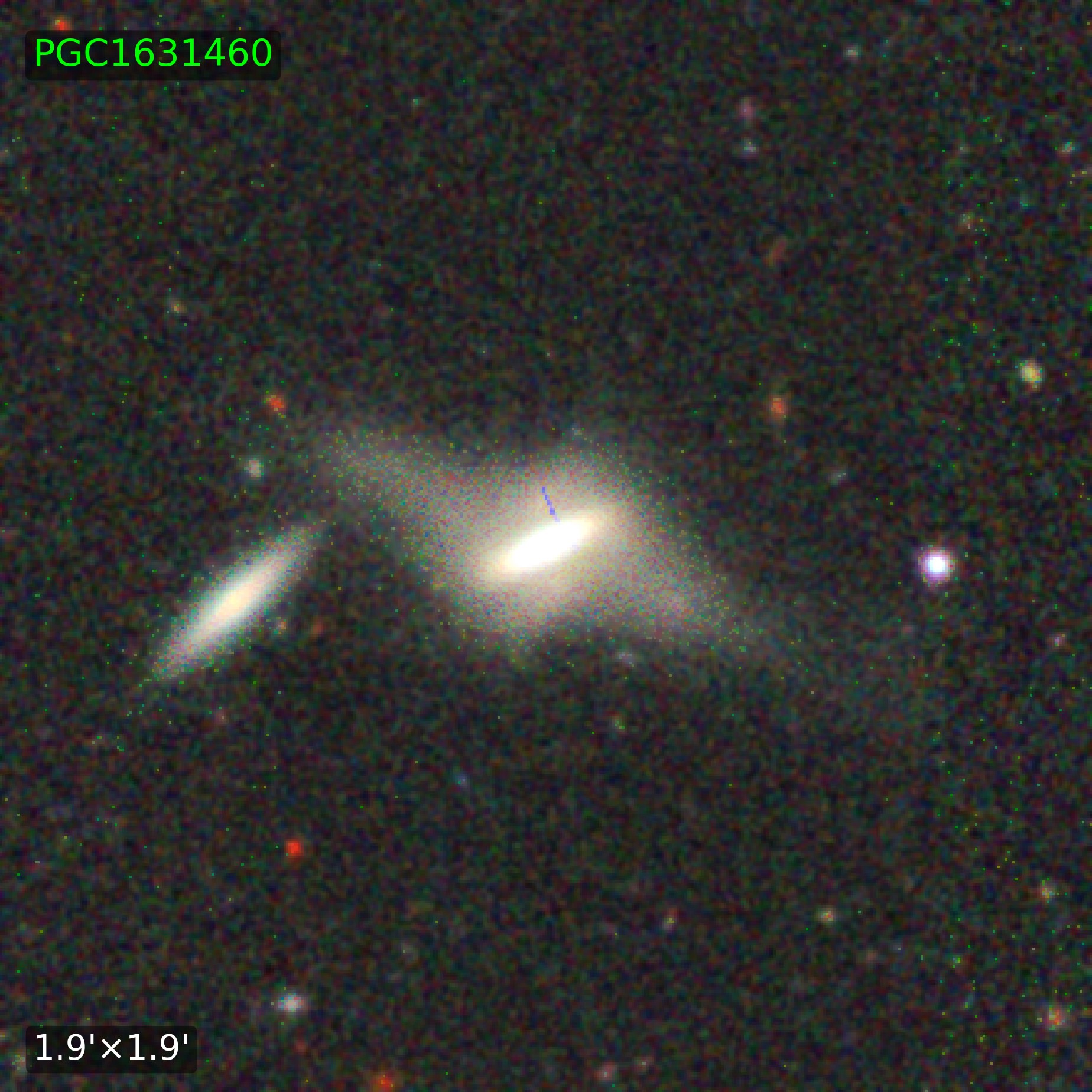}
    \end{subfigure}
    
    \caption{Examples of galaxies with multiple polar structures from our catalog.}
    \label{fig:multiple_structures}
\end{figure}

After refining our sample to include only the highest-quality PSG candidates, we further subdivided the galaxies into six PSG subtypes (denoted as \texttt{PSG\_TYPE\_1} in the catalog table described in Appendix~A), following the classification criteria outlined in Sect.~\ref{subsec:nom}. However, for approximately 27\% of the PSG candidates, a single classification was insufficient to capture the complexity of their morphologies. Because some galaxies exhibit multiple polar structures within a single system, we assigned each PSG a secondary subtype (\texttt{PSG\_TYPE\_2} in the same catalog table) using the same six principal categories employed in the primary classification. For instance, a galaxy classified as a PR subtype with a secondary subtype of PTS exhibits both a symmetric PR and polar tidal structures, the latter likely associated with its formation history. To decide which structure would determine the primary versus secondary classification, we considered the prominence of each structure based on size and brightness. Also, fully formed polar structures were always prioritized over PTSs, since tidal structures are more transient in nature. This additional classification, where applicable, will be particularly valuable for our forthcoming study of LSB features surrounding PSGs. Representative examples of galaxies with multiple polar structures are shown in Fig.~\ref{fig:multiple_structures}.

In addition to classifying the polar structures themselves, we also performed an independent morphological classification of the PSG host galaxies. Because PSGs are often morphologically peculiar systems, the classifications provided in public databases such as NED\footnote{\url{https://ned.ipac.caltech.edu/}} and HyperLeda\footnote{\url{http://atlas.obs-hp.fr/hyperleda/}} \citep{2014A&A...570A..13M} cannot be considered fully reliable for our purposes. Therefore, we visually reclassified all host galaxies into six broad morphological categories: elliptical (E), lenticular (S0), S0–a, spiral (S), irregular (Irr), and indeterminate (M). The latter category includes systems whose host morphology could not be confidently determined, typically due to ongoing or recent merging activity. Galaxies were classified as S0–a when the disk morphology could not be visually distinguished between a lenticular and a spiral type. A comparison between our morphological classifications and those listed in HyperLeda is discussed in Sect.~\ref{subsec:host_class}.

\subsection{Data preparation and photometric analysis}\label{subsec:phot_analysis}

For all selected candidates, we followed a standard data-preparation procedure consisting of the following main steps: (1) background correction; (2) masking of foreground stars, background and foreground galaxies, and image artifacts; (3) refinement of the galaxy center; (4) isophotal analysis using the \texttt{IRAF ELLIPSE/FITTING} task; and (5) measurement of the integrated photometric properties of each galaxy.

Since the imaging data we used were taken from the DESI Legacy~DR10, no additional large-scale background subtraction was required, as the survey pipeline already provides well-calibrated, sky-subtracted images and these were sufficient for our photometric analysis. Nevertheless, we verified the background quality for each frame by measuring the median and standard deviation of the sky level in multiple source-free regions. These values were used to confirm the uniformity of the background and to estimate the photometric depth of each image, ensuring consistency across all analyzed fields.

The masking of the background and foreground sources was performed using the \texttt{MTObjects} package\footnote{\url{https://github.com/CarolineHaigh/mtobjects}}, which implements a multi-thresholding segmentation algorithm based on the original C implementation by \citet{teeninga2015improved}. This method efficiently identifies and masks stars, galaxies, and other contaminating sources in the vicinity of each target. All automatically generated masks were subsequently inspected and refined manually to ensure that no parts of the PSGs or their surrounding LSB features were inadvertently masked out.

We also redetermined the centers of all galaxies, as for some sources the catalog coordinates were originally based on poorly resolved 2MASS \citep{2006AJ....131.1163S} or WISE \citep{2010AJ....140.1868W} observations. This step was also necessary because PSGs are morphologically peculiar systems, and their centers must be measured in a consistent and uniform manner across the entire sample. The galaxy centers were measured directly from the DESI Legacy imaging using the following procedure. A small square cutout, scaled according to the galaxy’s extent, was extracted around the nominal galaxy center and the previously created mask was applied to exclude bad pixels. The initial galaxy coordinates served as the starting point for a parametric refinement, in which a 2D Gaussian model was fitted to the cutout using a Levenberg–Marquardt least-squares optimizer, with masked regions excluded through pixel weighting. When the optional internal masking was enabled, residuals between the data and the best-fit model were used to iteratively construct an internal mask of outliers above a specified threshold. The fitting procedure was repeated several times to ensure stability. After convergence, the fitted center was transformed back to full-frame coordinates, converted to sky coordinates (RA, Dec) using the image world coordinate system, and compared with the image midpoint. With respect to the resulting refined centers , if they differed by more than one pixel from the input coordinates, they were adopted as the reference positions for all subsequent photometric measurements.

We used the \texttt{IRAF ELLIPSE/FITTING} task \citep{1987MNRAS.226..747J} to derive azimuthally averaged surface-brightness profiles and cumulative (growth) curves for each galaxy in the $g$, $r$ and $z$ wavebands. From the growth curves, we measured (per filter) the flux enclosed by the $\mu=26~\mathrm{mag~arcsec^{-2}}$ isophote and the corresponding isophotal radius $R_{26}$, obtained by solving for the outermost radius at which the spline representation of $\mu(R)$ equals $26~\mathrm{mag~arcsec^{-2}}$. 

The uncertainties in the magnitudes and isophotal radii were estimated through a Monte Carlo procedure that propagated three primary sources of error: (1) per–isophote intensity uncertainties from the \texttt{ELLIPSE} output; (2) global sky–background fluctuations modeled as a random additive offset to all isophotes; and (3) the adopted photometric zeropoint uncertainty taken from \citep{2023RNAAS...7..105Z}. For each galaxy, 1000 random realizations of its surface–brightness profile were generated by perturbing these quantities, and the apparent magnitude and $R_{26}$ were recomputed in each iteration. The standard deviation of the resulting distributions was taken as the statistical uncertainty, while the zero-point term was added in quadrature to obtain the total error budget.

All selected PSG candidates were cross-matched with the SGA, HyperLeda, and NED databases to obtain the available photometric and morphological information for each galaxy. In addition, all galaxies were cross-matched with the SDSS~DR19 database \citep[][submitted to AAS Journals]{2025arXiv250707093S} to retrieve spectroscopic redshifts. They were then cross-matched with the DESI Legacy~DR9 and DR10 releases to obtain the photometric redshift estimates. For galaxies with available spectroscopic redshifts ($z_\mathrm{sp}$), we adopted these values as their redshifts, $z$. When spectroscopic measurements were unavailable, we instead used the corresponding photometric redshifts ($z_\mathrm{ph}$). To ensure that there are no systematic deviations between the two, we compared DESI Legacy~DR10 photometric redshifts with SDSS spectroscopic redshifts for 1,034 galaxies with both measurements. This comparison yields a linear relation of $z_\mathrm{ph} = 1.015\,z_\mathrm{sp}$, with a coefficient of determination of $\rho^2 = 0.932$ and a median relative error of $|z_\mathrm{sp} - z_\mathrm{ph}| / z_\mathrm{sp} = 0.161$. 

Luminosity and angular-diameter distances were then calculated using the $z$ redshifts, assuming a standard flat $\Lambda$CDM cosmology based on \textit{Planck} results \citep{2020A&A...641A...6P,2024A&A...682A..37T}, with $H_0 = 67.4~\mathrm{km\,s^{-1}\,Mpc^{-1}}$, $\Omega_\mathrm{M} = 0.315$, and $\Omega_\Lambda = 0.685$. For all objects, we also computed the Galactic extinction in the $g$, $r$, and $z$ bands using the \citet{1998ApJ...500..525S} reddening map and the conversion coefficients listed in Table~6 of \citet{2011ApJ...737..103S}. $K$-corrections were applied using the publicly available code described by \citet{2010MNRAS.405.1409C} and \citet{2012MNRAS.419.1727C}, while the resulting extinction- and $K$-corrected absolute magnitudes were computed using the \texttt{IRAF} apparent magnitudes. Throughout this paper, all magnitudes are given in the AB photometric system.

\section{Catalog description}\label{sec:cat_descr}

The structure of the catalog table is described in detail in Appendix~\ref{app:table}. The catalog table is available at the CDS\footnote{\url{https://cdsarc.cds.unistra.fr/viz-bin/qcat?J/A+A/}}. The accompanying atlas, a PDF containing enhanced RGB images of all galaxies, is available at Zenodo\footnote{\url{https://zenodo.org/}}.

\begin{figure*}
    \centering
\includegraphics[width=0.98\linewidth]{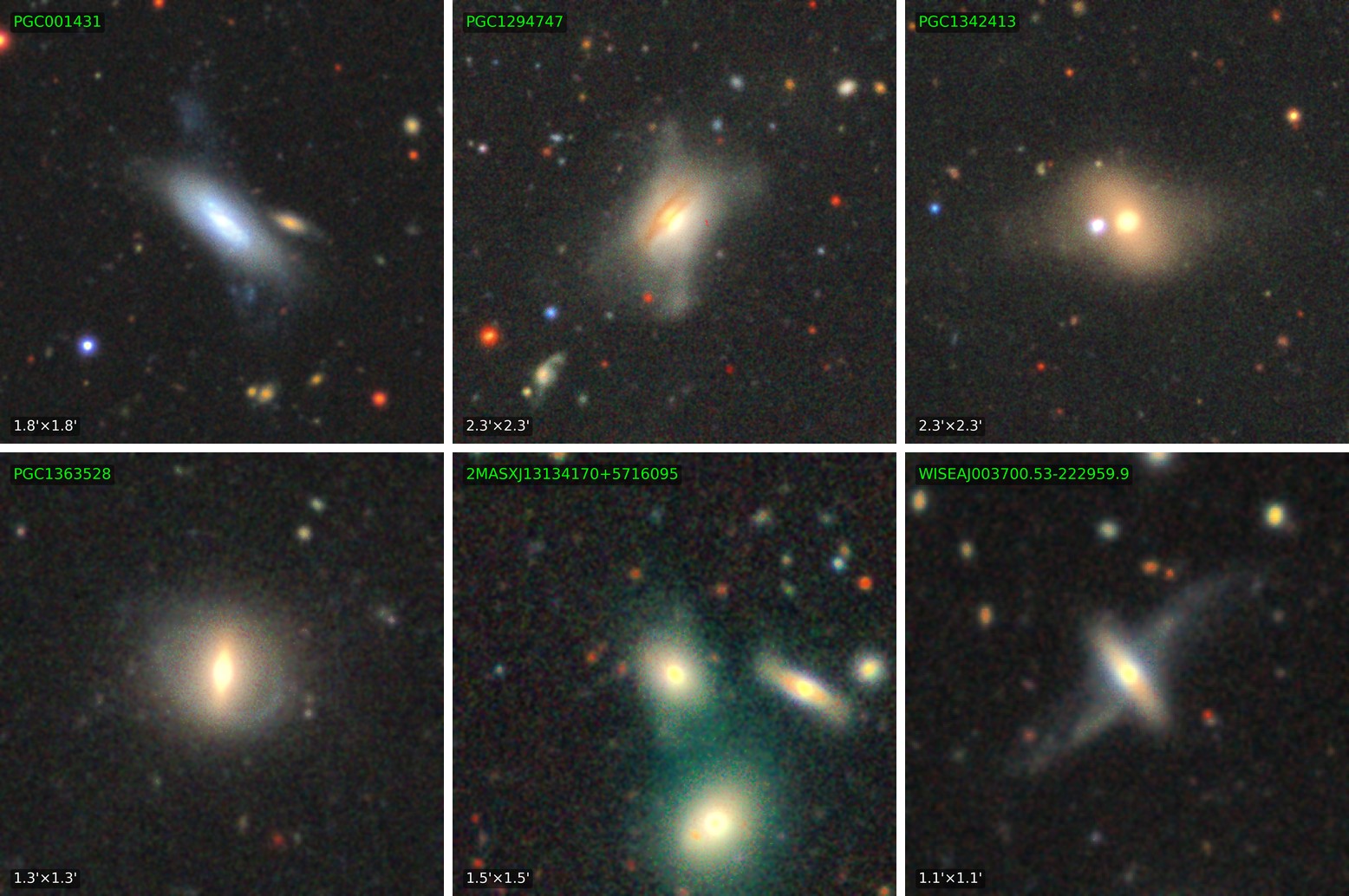}
\includegraphics[width=0.98\linewidth]{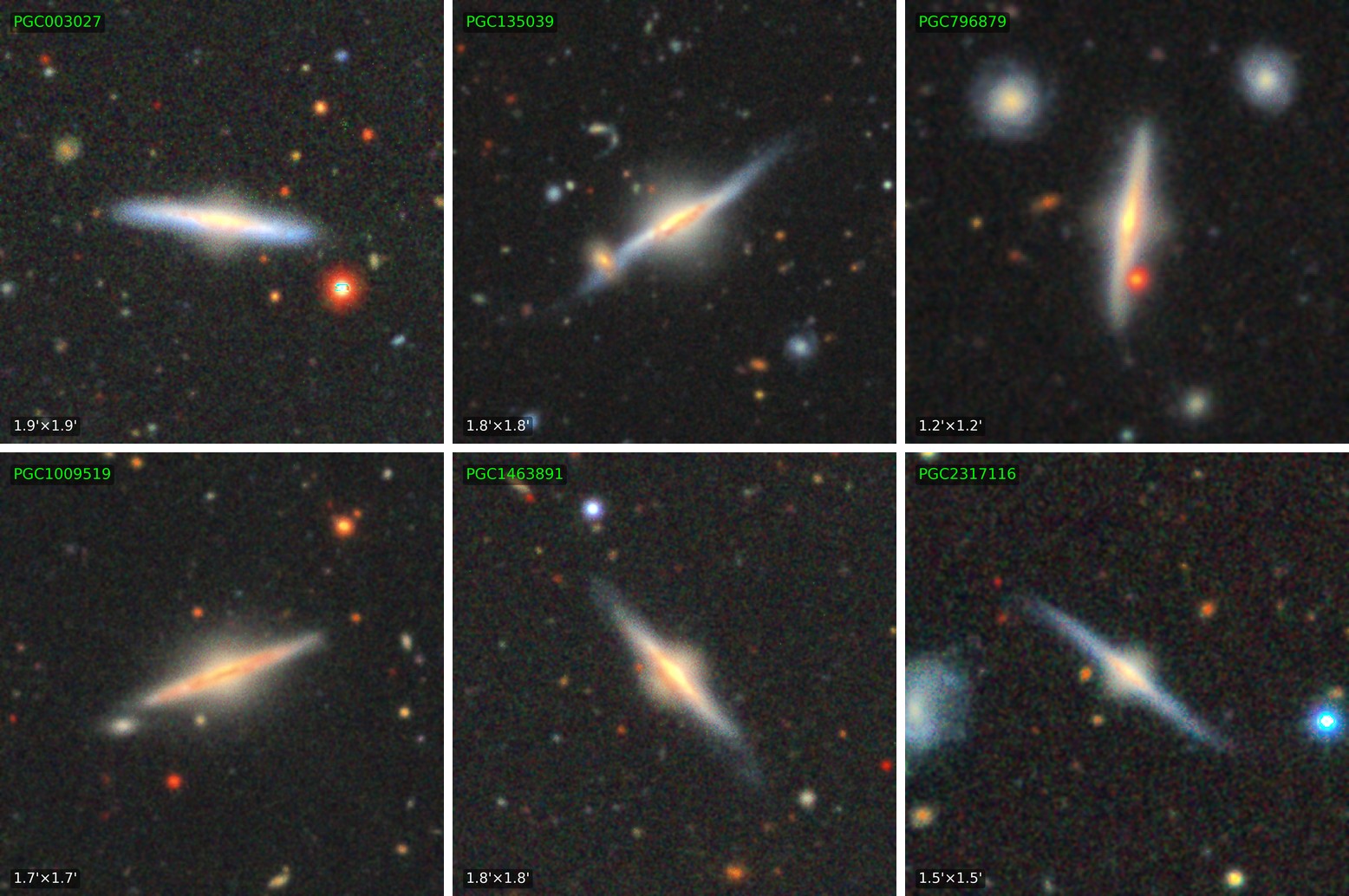}
    \caption{Randomly selected examples of PRGs (top two rows) and PBs (bottom two rows) from our catalog.}
    \label{fig:prg_pb_examples}
\end{figure*}

\begin{figure*}
    \centering
\includegraphics[width=0.98\linewidth]{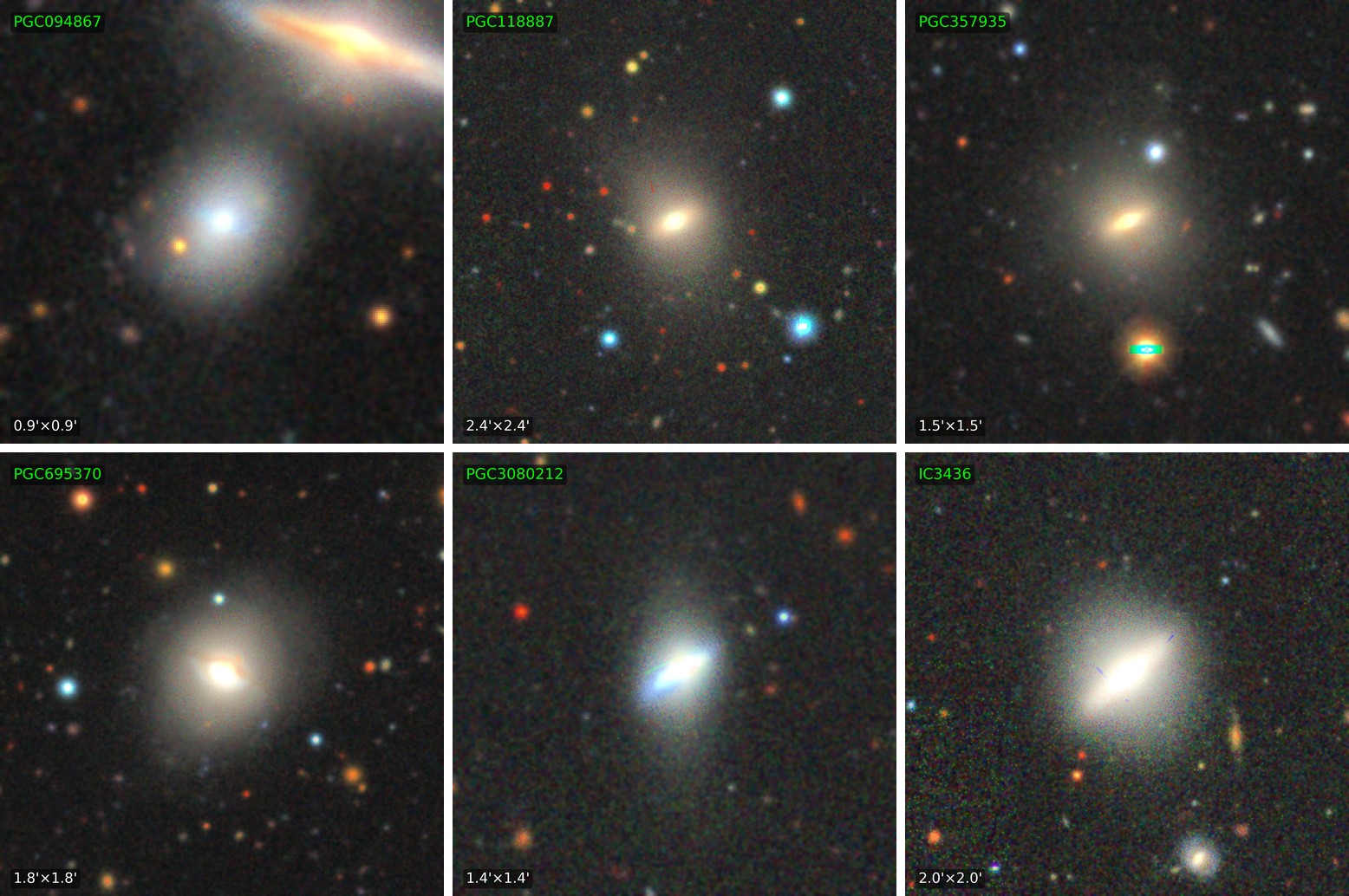}
\includegraphics[width=0.98\linewidth]{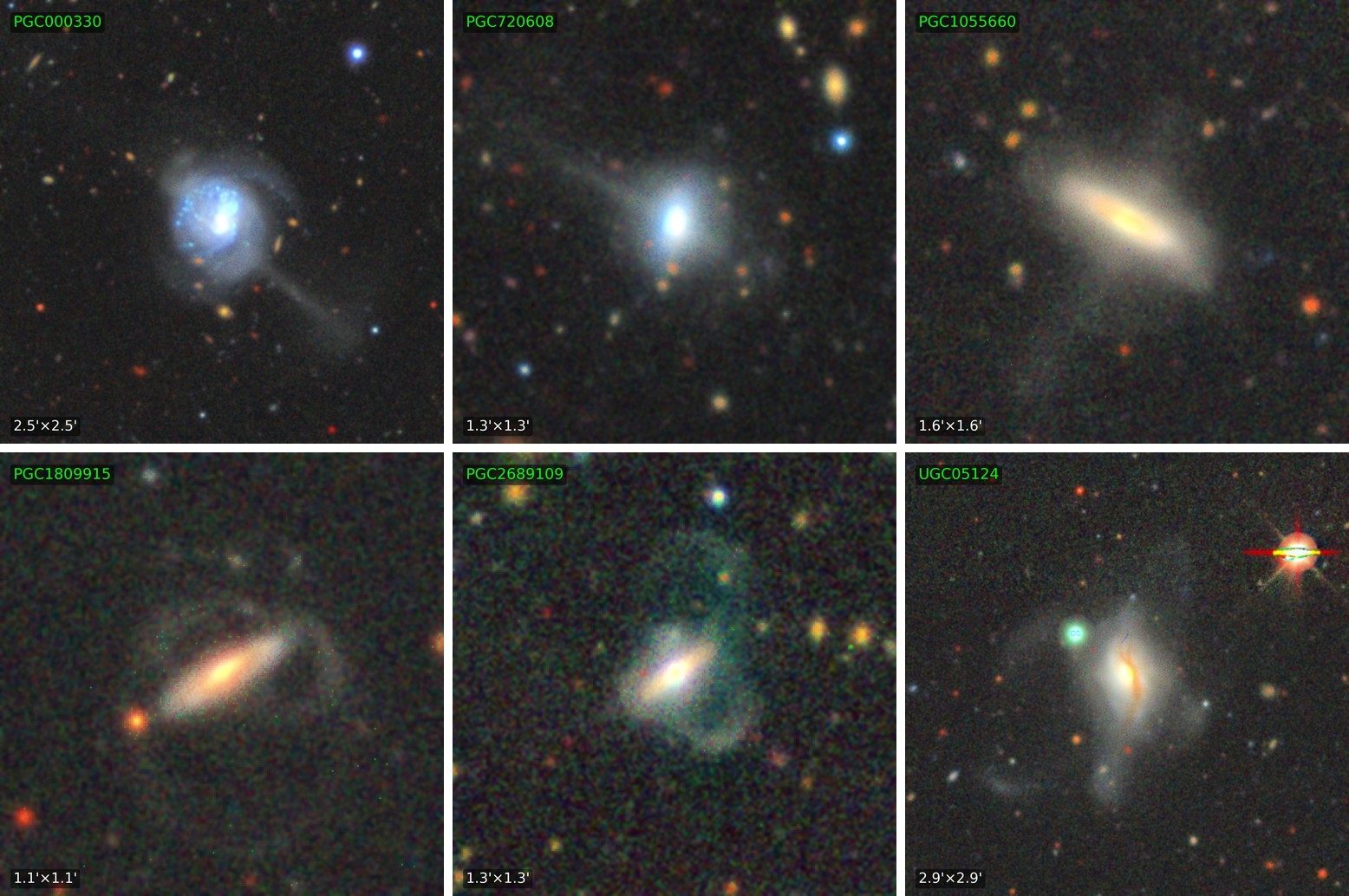}
    \caption{Randomly selected examples of PHs (top two rows) and PTSs (bottom two rows) from our catalog.}
    \label{fig:ph_pts_examples}
\end{figure*}

The final PSG candidate catalog, designated as the Catalog of Unusual Galaxies with polar Structures in the DESI Legacy Imaging Surveys (COUGS-DESI), comprises 2,989 PSGs from the DESI Legacy~DR10. It contains both newly identified candidates and strong candidates previously reported in the literature. Of the 2,989 galaxies in our catalog, 2,626 ($\sim88\%$) have been included in the SGA. Each galaxy in the catalog is assigned to one or more subtypes based on its visual morphology in the DESI Legacy~DR10 photometric data. The catalog includes six principal subtype categories: PRs, PBs, PHs, PDLs, PTSs, and other polar structures, as described in Sect.~\ref{sec:cat_comp}. Representative examples of PRs, PBs, PHs, and PTSs are presented in Figs.~\ref{fig:prg_pb_examples}--\ref{fig:ph_pts_examples}.

\begin{figure}
    \centering
    \includegraphics[width=1\linewidth]{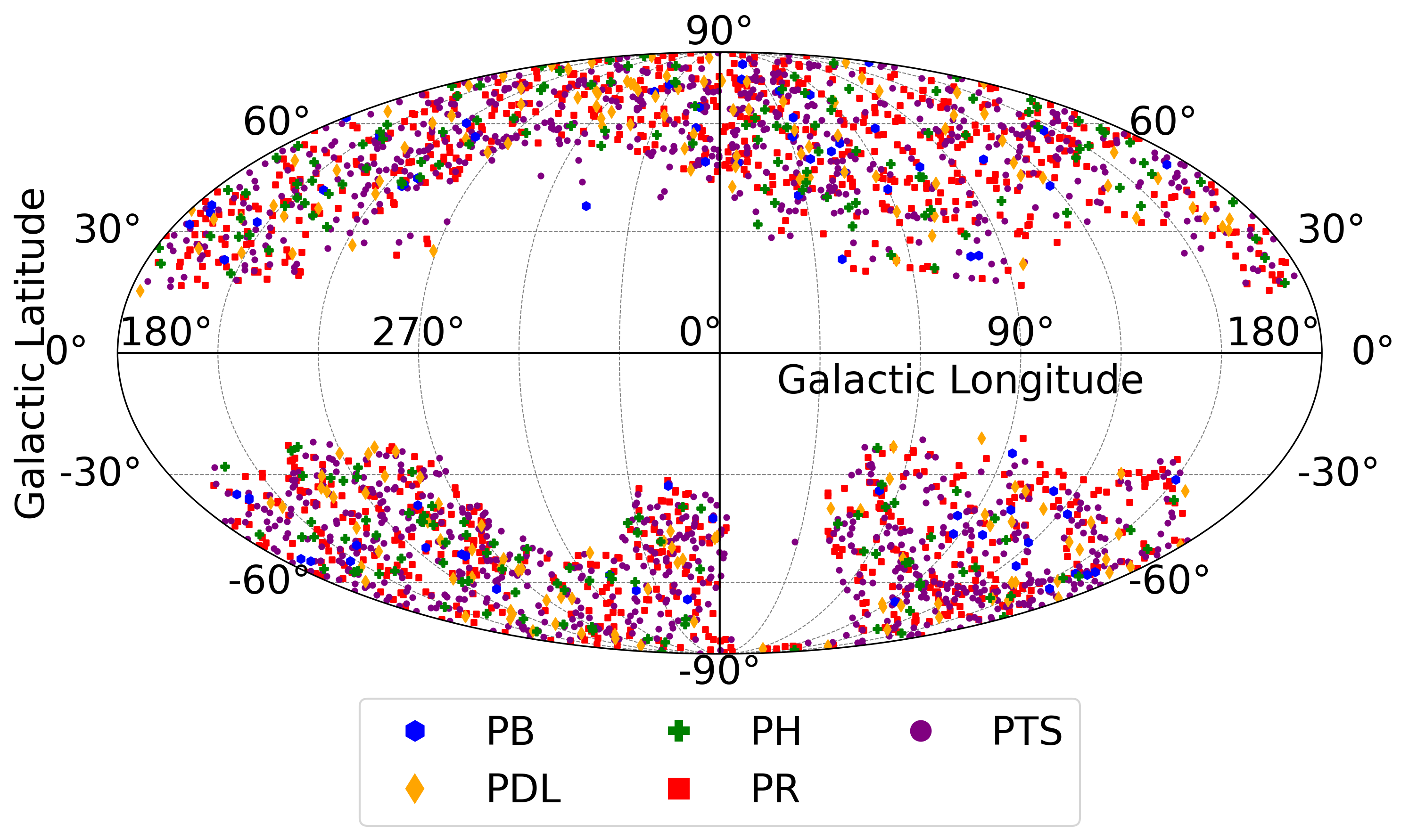}
    \caption{Distribution of all major PSG types across the sky, displayed in Galactic coordinates (longitude and latitude).}
    \label{fig:PSG_Distribution}
\end{figure}

In Fig.~\ref{fig:PSG_Distribution}, we present the distribution of PSGs in our catalog across the celestial sphere, shown in Galactic coordinates. As expected, PSGs are uniformly distributed over the sky, with no significant clustering or concentration toward any particular Galactic region. The different PSG types also exhibit a similarly uniform distribution, indicating that no specific class is preferentially located in any part of the sky. This overall uniformity suggests that our sample is not strongly affected by selection biases.

\subsection{General statistics of the sample}\label{subsec:stat}

The expanded sample of PSGs provides an unprecedented opportunity for a statistical study of these systems. Below we present the general statistics of PSGs in our catalog.

The counts and percentages of catalog galaxies assigned to each morphological class (PTSs, PRs, PDLs, PHs, PBs, and other polar structures) are summarized in Table~\ref{tab:psg_types}. It should be noted that the sample of "other" polar structures is significantly incomplete, as these systems are not the primary focus of our catalog. The results of both the primary and secondary morphological classifications indicate that PTSs are particularly common in our catalog, appearing in more than half of the galaxies, either as their primary (\texttt{PSG\_TYPE\_1}) or secondary (\texttt{PSG\_TYPE\_2}) classification. This prevalence is not unexpected, given that tidal structures and the minor merger events responsible for their formation are nearly ubiquitous in the Local Universe \citep{2010AJ....140..962M, 2023A&A...671A.141M, 2025A&A...700A.176M}.

\begin{table}
\centering
\caption{Counts ($N$) and percentages ($p$) of galaxies in each PSG type.}
\label{tab:psg_types}
\begin{tabular}{lrrrr}
\hline
Category & $N_1$ & $p_1$ [\%] & $N_2$ & $p_2$ [\%] \\
\hline
\hline
PTS & 1315 & 44.0 & 645 & 79.8 \\
PR & 1113 & 37.2 & 23 & 2.8 \\
PDL & 185 & 6.2 & 117 & 14.5 \\
PH & 216 & 7.2 & 5 & 0.6 \\
PB & 75 & 2.5 & 0 & 0.0 \\
Other & 85 & 2.8 & 18 & 2.2 \\
\hline
\end{tabular}
\tablefoot{The second and third columns correspond to the \texttt{PSG\_TYPE\_1} classification. The fourth and fifth columns correspond to the \texttt{PSG\_TYPE\_2} classification (if any).}
\end{table}

After PTSs, the most common PSG subtype in the catalog is PRs, which account for roughly one-third of all systems. PDLs and PHs constitute the next most frequent categories. As discussed in Sect.~\ref{subsec:nom}, PDLs may correspond to gaseous PRs that lack a detectable stellar component, or cases in which the stellar ring is strongly obscured by dust. In contrast, PHs could represent faint, highly diffuse stellar counterparts or unresolved PRs. The depth of the observations strongly influences our ability to identify polar halos; with deeper imaging, their relative fraction is expected to increase \citep{2021MNRAS.506.5030M}. Improved spatial resolution should likewise reveal a larger number of inner polar structures than reported here. Moreover, deep H{\sc I} observations may uncover an even higher fraction of galaxies hosting gaseous polar structures \citep{2023MNRAS.525.4663D}. In contrast, PBs appear to be the rarest type of polar structure in our catalog, most likely due to their intrinsically low surface brightness and the narrow range of host-galaxy inclinations under which they can be identified from photometric data alone.

\begin{figure}
    \centering
    \includegraphics[width=\linewidth]{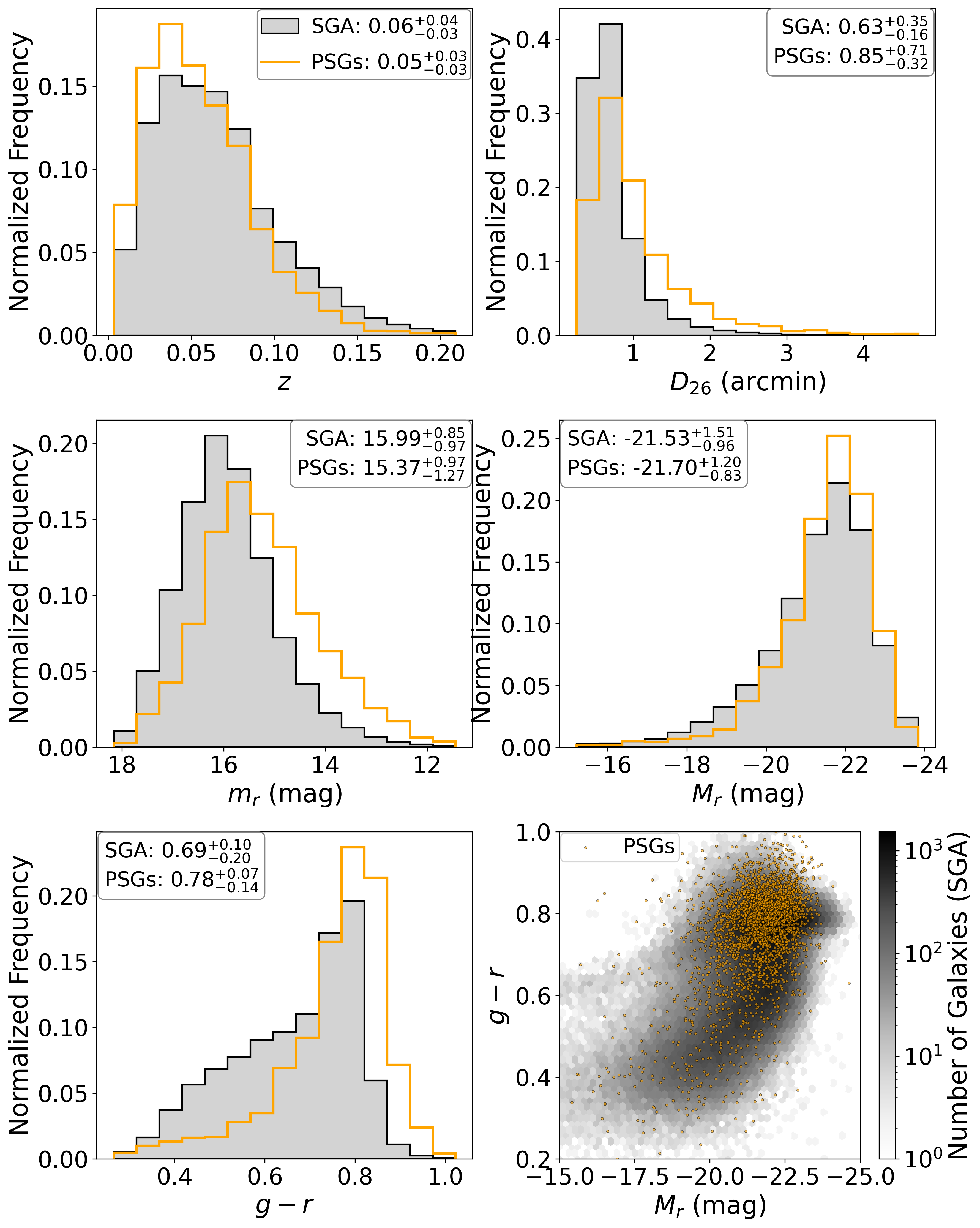}
\caption{Histograms of key parameters for the full SGA sample and the PSG subsample, with outliers (less than 1\%) removed. Each panel lists the median value, with superscripts and subscripts denoting the offsets to the 84th and 16th percentiles (i.e., median $\pm \Delta$). The bottom-right panel presents the color–magnitude diagram, where the SGA distribution is shown as a density map and the PSGs are overplotted as orange points.}
    \label{fig:gen_stats_zoomed}
\end{figure}

We compared the full SGA sample with our subsample of PSGs to identify statistical trends within our catalog. For this analysis, we used galaxy parameters provided by the SGA \citep{2023ApJS..269....3M}, including the integrated $r$-band magnitude measured within the 26~mag\,arcsec$^{-2}$ isophote, redshift (primarily photometric, with spectroscopic values when available), absolute $r$-band magnitude (corrected for Galactic extinction and $K$-correction), $g-r$ color, and the $D_{26}$ isophotal diameter\footnote{See \url{https://www.legacysurvey.org/sga/sga2020/} for details.}. For each parameter, we excluded outliers (constituting less than 1\% of the total number of galaxies in each sample) to better highlight the overall trends in the data. The resulting distributions for these parameters, along with a color–magnitude diagram, are shown in Fig.~\ref{fig:gen_stats_zoomed}.

We also conducted a two-sample Kolmogorov-Smirnov (K-S) test for the PSG and SGA galaxy distributions. Using parameter values from the SGA, we compared the empirical cumulative distribution functions (CDFs) of the entire SGA and our selected PSGs for apparent $r$-band magnitude, redshift, absolute $r$-band magnitude, $g-r$ color, and $D_{26}$ isophotal diameter. The K--S test results confirm the presence of two distinct populations at the $\alpha = 0.001$ significance level in every case: the K--S statistic spans $D = 0.087$--0.294, indicating modest to moderately strong differences between the cumulative distributions of the two samples, and all p-values are $\le 1.9\times10^{-19}$. The strongest similarity between the whole SGA and PSG CDFs was for absolute magnitude and redshift. The CDFs were more strongly differentiated when compared using apparent magnitude, $D_{26}$ isophotal diameter, and $g-r$ color. Since the K--S test alone does not explain the reasons for the observed CDFs, we discuss likely explanations for the observed similarities and differences in the different CDFs below, including both selection effects and intrinsic galactic properties.

Compared to the overall SGA population, PSGs tend to have slightly lower redshifts. This is an observational bias arising from the greater ease of identifying faint polar structures at low redshifts than at higher ones. As distance increases, the structural features of galaxies become progressively less apparent due to limited angular resolution and surface-brightness sensitivity. This is a limitation that affects the detection of polar structures and other galaxy structural features equally. Consequently, although relatively few PSGs have been identified at $z > 0.1$, this likely does not reflect their true abundance at higher redshifts (see our discussion of future prospects for PSG studies in Sect.~\ref{subsec:future}); rather, many such systems may simply remain undetectable with ground-based observations.

One of the most notable differences between the PSGs and the full SGA sample is observed in their apparent magnitudes. On average, PSGs appear significantly brighter ($m_r = 15.37$) than the general SGA population ($m_r = 15.99$). This difference primarily reflects the previously discussed redshift distribution of the PSGs, as well as the fact that the observed PSGs are predominantly giant and intermediate-mass galaxies, as indicated by their absolute magnitude distribution.

The $D_{26}$ isophotal diameters are, on average, larger for the PSGs ($D_{26} = 0.85$~arcmin) than for the full SGA sample ($D_{26} = 0.63$~arcmin). This difference likely reflects the same selection bias, as polar structures are more discernible in galaxies that appear visually more extended. In addition, the larger angular diameters of PSGs may result from the extra light contributed by the polar components, which can extend well beyond the main body of the host galaxy \citep{2019MNRAS.483.1470R} and increase the overall isophotal radius. In many cases, polar rings or disks add measurable surface brightness at larger galactocentric distances, effectively expanding the region encompassed by the $D_{26}$ isophote. Consequently, the presence of a polar structure not only modifies the morphological appearance of these systems but also influences their general characteristics.

The $g-r$ color distribution also reveals a clear discrepancy between the PSGs and the full SGA sample. Compared to the full SGA, PSGs show an overabundance of redder systems relative to bluer ones. This trend is expected, as the $g-r$ color is predominantly determined by the light from the bright host galaxy, and many polar structures are associated with lenticular or elliptical hosts (see Sect.~\ref{subsec:host_class}), which are typically red and quiescent. Although many PSGs are multicolored (e.g., a PRG may consist of a red elliptical or lenticular host surrounded by a blue, star-forming ring or disk),the host galaxy generally dominates the total light, producing a skew toward redder $g-r$ colors. While some PSG subtypes, particularly PBs, exhibit blue host galaxies, the overall distribution of PSG types favors systems with red hosts. Conversely, although many polar structures themselves are blue owing to ongoing star formation, others persist long enough to redden as star formation ceases and stellar populations evolve (see, e.g., Fig.~\ref{fig:prg_pb_examples} for examples of both blue and red PRs and hosts, and the discussion in \citealt{2025PASA...42...56A}).

In the full SGA, two well-defined populations are evident: the blue cloud and the red sequence, with the green valley lying between them \citep[e.g.,][]{2004ApJ...600..681B,2007ApJS..173..342M}. These distinct groupings reflect the bimodal nature of the galaxy population, corresponding to star-forming, quiescent, and transitional systems, respectively. From the color-magnitude diagram in Fig.~\ref{fig:gen_stats_zoomed}, we see that PSGs follow the general trend of the SGA galaxies, though skewed toward the red sequence \citep[see also][]{2024A&A...681L..15M}. Additionally, our PSG sample has a somewhat higher proportion of late-type galaxies compared to existing PRG catalogs. In COUGS-DESI, 53.6\% of PSG candidates have spiral hosts (even when S0 and S0-a galaxies are combined). By comparison, for the galaxies from the literature included in our catalog, the spiral fractions are 50\% in the PRC, 48.7\% in the SPRC, 36\% in the catalog of \citet{2019MNRAS.483.1470R}, and 45.6\% in the catalog of \citet{2024A&A...681L..15M}. This is discussed further in Sect.~\ref{subsec:host_class}.

\begin{figure}
    \centering
    \includegraphics[width=\linewidth]{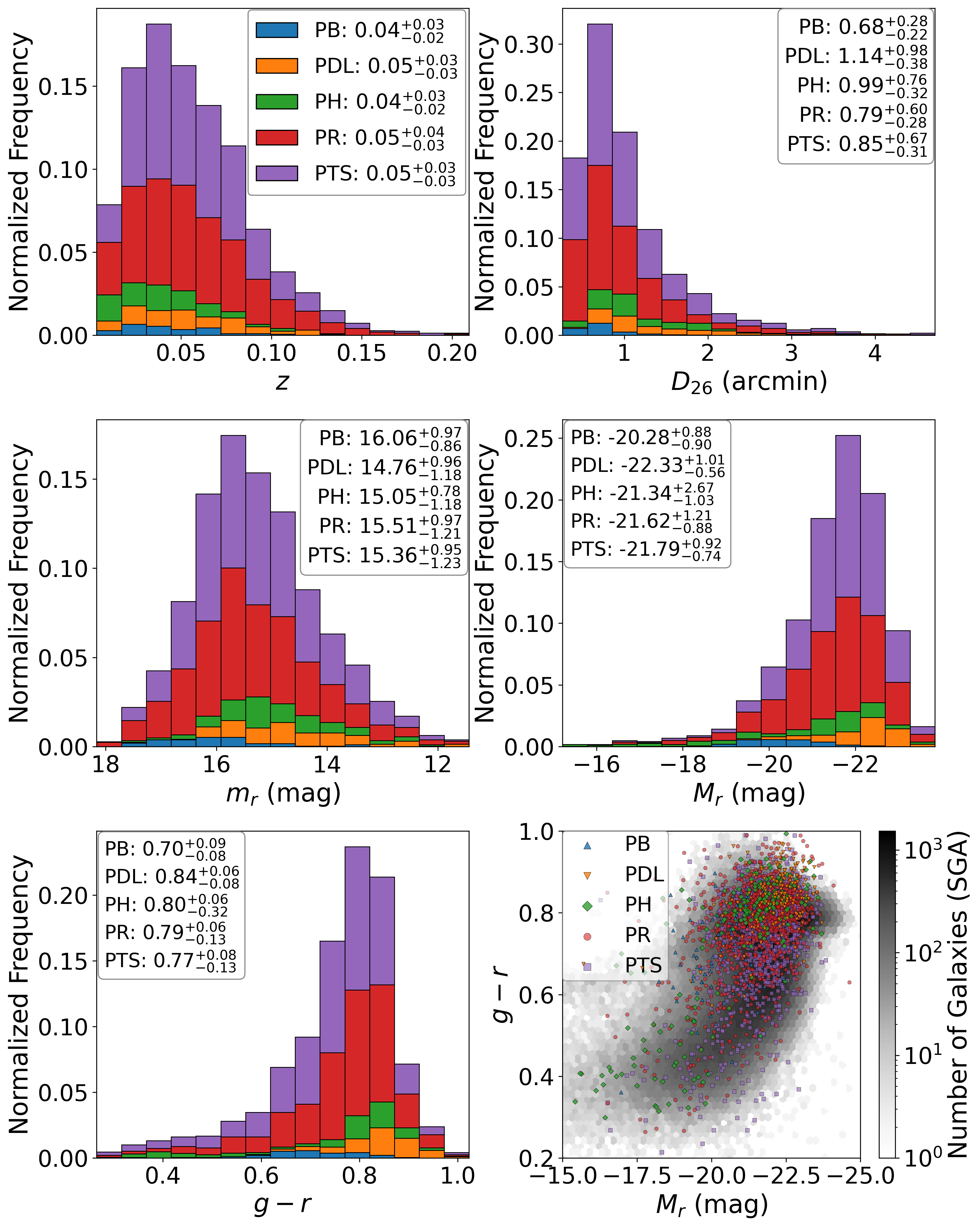}
\caption{Same as Fig.~\ref{fig:gen_stats_zoomed}, but shown separately for the different subtypes (\texttt{PSG\_TYPE\_1}) of polar structures. The stacked format emphasizes the relative contribution of each PSG subtype.}
\label{fig:gen_stats_zoomed_subtypes}
\end{figure}

In addition to comparing the properties of the full PSG sample with those of the SGA, it is also instructive to examine differences among the individual PSG subtypes. As in Fig.~\ref{fig:gen_stats_zoomed}, Fig.~\ref{fig:gen_stats_zoomed_subtypes} presents histograms of redshift, isophotal diameter, apparent and absolute magnitudes, and $g-r$ color, as well as the color--magnitude diagram, but in this figure, they are separated into the five PSG subtypes.

We also performed two-sample Kolmogorov--Smirnov (K-S) tests for the distributions of each parameter for each PSG subtype. In the majority of cases, the K-S test results indicate statistically distinct populations at the $\alpha = 0.001$ level. However, there are several exceptions, where the K-S statistic, $D$, shows that the populations are quite similar. For example, when comparing the redshift distributions between subtypes, two broad groupings emerge: the redshift distributions of PRs, PTSs, and PDLs cannot be reliably distinguished from one another ($D_{z,~PR-PDL}=0.05$, $D_{z,~PR-PTS}=0.04$, $D_{z,~PDL-PTS}=0.08$), while the PB and PH redshift distributions show only marginal differences ($D=0.17$ with $p = 0.063$, i.e. they are distinct only at the $\alpha \approx 0.1$ level). The PB and PH redshift populations are more similar to each other than to the redshift distributions of PRs, PTSs, and PDLs---and vice versa. The other K-S test that yielded only a marginal difference between the two populations was for the $m_r$ distributions for PDLs and PHs ($D=0.14$, $p=0.04$). For all other parameter--subtype comparisons where the K--S test indicated distinct populations at our adopted significance threshold, we found $p \le 0.001$.

The parameters for which the K-S test most reliably showed distinct populations for each PSG subtype were absolute magnitude $M_r$ and $g-r$ color. When considering either of these parameters, the K-S test statistic, $D$, ranges from 0.1--0.73 with corresponding $2.5\times10^{-28} \le p \le 1\times10^{-5}$ (in the case of $M_r$) or 0.1--0.61 with corresponding $7.9\times10^{-19} \le p \le 2\times10^{-5}$ (in the case of $g-r$ color). This confirms the separate populations at the $\alpha=0.001$ level for all combinations of PSG subtypes. Performing the K-S test using the parameters of apparent magnitude $m_r$ and $D_{26}$ isophotal diameter also yields separate populations for all combinations of PSG subtypes, but with higher $\alpha$ levels (ranging from 0.005 up to 0.05) in some cases.

As shown in Fig.~\ref{fig:gen_stats_zoomed_subtypes}, PB galaxies are, on average, intrinsically fainter than other PSG subtypes. Their lower luminosity is one of the characteristics that distinguishes PBs from other types of polar structures in our catalog. PBs also tend to exhibit bluer $g-r$ colors compared to other PSG subtypes, reflecting the predominance of spiral hosts within this class. Indeed, most PB hosts are actively star-forming spiral galaxies, as evidenced by their position in the color–magnitude diagram.

Galaxies with PDLs are, on average, significantly brighter than other PSG subtypes in both apparent and absolute magnitude. PDLs are typically observed in elliptical galaxies \citep{1987IAUS..127..135B}. Because PDLs are seen in absorption against the stellar background of the host, they are most easily detected in bright, featureless, and red galaxies, which also tend to exhibit slightly redder $g-r$ colors. PDL galaxies generally have larger angular sizes than other PSG subtypes, a consequence of the fact that in angularly smaller elliptical systems, a polar dust lane would be smeared out and thus rendered undetectable. Owing to their red colors and high luminosities, PDL galaxies occupy the top-right region of the color–magnitude diagram, relative to other PSGs.

Among PSG subtypes, PHs uniquely demonstrate a bimodal $g-r$ color distribution, with a red and blue peak. Since polar halo structures are typically very faint, most of the contribution to the color of galaxies with PHs comes from the host galaxy. The two peaks of the color distribution for PHs represent the two primary host types for these galaxies. The red peak corresponds to lenticular and quiescent spiral host galaxies, whereas the blue peak corresponds to star-forming hosts.

\subsection{Host classification} \label{subsec:host_class}

\begin{figure*}
    \centering
    \includegraphics[width=1\linewidth]{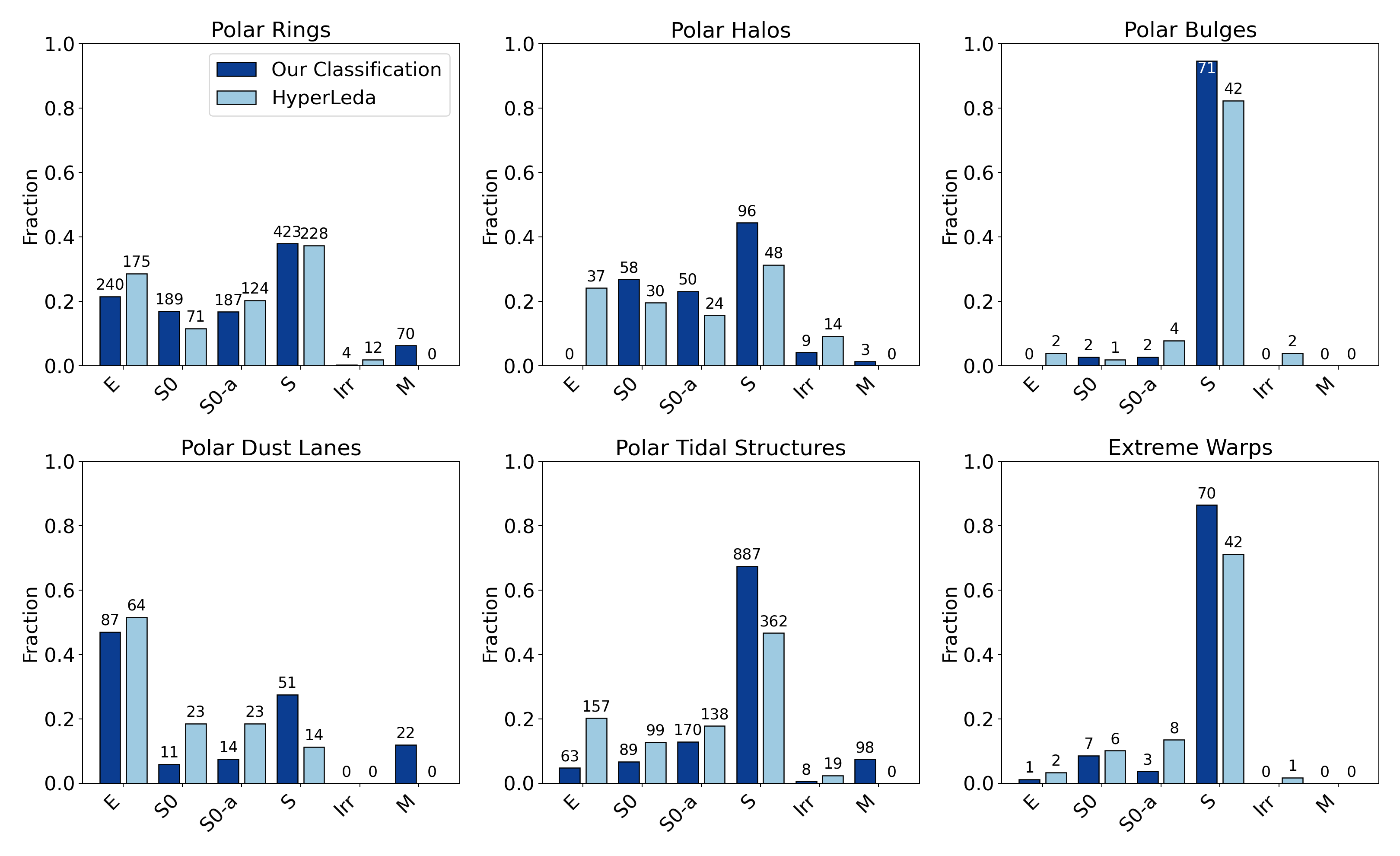}
    \caption{Host-galaxy classification results for the five PSG categories and the extremely warped galaxies in our catalog. A comparison with the HyperLeda classification is also shown. The numbers above the columns indicate the total number of objects of each morphological type within each PSG category.}
    \label{fig:hostbarcharts}
\end{figure*}

We present the statistical properties of the host-galaxy classifications for our catalog. The results of the host morphology (see the \texttt{MORPHTYPE} column in the catalog table) are shown in Fig.~\ref{fig:hostbarcharts}, and the PSG fractions for each host type in the SGA are provided in Table~\ref{tab:psg_fractions_threeparts}.

\begin{table}
\centering
\caption{Fraction of each PSG subtype per host galaxy morphological type as percentages.}
\begin{tabular}{l|c|cccc}
\hline
Category & $N_{\mathrm{PSG}}$ & E & S0 & S & M \\
\hline
\hline
PR & 1103 & 0.41\% & 0.95\% & 0.15\% & 1.09\% \\
PH & 206 & 0.00\% & 0.31\% & 0.03\% & 0.04\% \\
PB & 69 & 0.00\% & 0.01\% & 0.02\% & 0.00\% \\
PTS & 1176 & 0.09\% & 0.66\% & 0.29\% & 1.17\% \\
\hline
TOTAL & 2554 & 69052 & 34526 & 272370 & 7672 \\
\hline
\end{tabular}
\tablefoot{The leftmost column gives the number of galaxies in each PSG subtype. PSG fractions are column-normalized by the total number of SGA galaxies in each morphological type, given in the last row. S0 and S0-a galaxies are combined into S0; S and Irr galaxies are combined into S. Galaxies classified as PDL are merged into the PR class.}
\label{tab:psg_fractions_threeparts}
\end{table}

For galaxies in our catalog that were also listed in the HyperLeda database, we compared our visual morphological classifications with those listed in HyperLeda  (1,769 objects). HyperLeda provides a detailed morphological classification scheme for spiral galaxies, including subtypes (e.g., Sa, Sb, Sc); whereas our classification is broader, grouping all spiral systems under a single category. Therefore, the comparison was restricted to the general morphological classes corresponding to our scheme. It should also be noted that HyperLeda does not include an M type (indeterminate due to merger); however, only 196 galaxies (6.5\% of the catalog) fall into this category, so this omission should not significantly affect the comparison. 

The results show that only 653 out of 1,769 galaxies (36.9\%) share the same general morphological type in both HyperLeda and our catalog. This relatively low level of agreement highlights the necessity of our independent, uniform reclassification of host morphologies for all PSGs in the catalog. The HyperLeda morphological classification is primarily drawn from heterogeneous literature sources---most notably the visually determined RC3 (Third Reference Catalogue of Bright Galaxies, \citealt{1991rc3..book.....D}) types based on photographic plates. However, our classification is based on uniform, high-quality digital imaging and consistent criteria, leading to systematic differences between the two.

In Fig.~\ref{fig:hostbarcharts}, we present the fractional distributions of host galaxy morphologies for various subtypes of polar structures and, for comparison, extreme warps. These fractions are shown for both our morphological classifications and those from the HyperLeda database. Despite the discrepancies between the two classification schemes, their overall trends are consistent. 

Interestingly, with the exception of PDLs, the majority of PSG host galaxies are spirals, as indicated by both classifications. PRs are observed across all morphological types, yet spiral hosts dominate (38\% in our catalog), even when S0 and S0/a galaxies are combined. This finding contrasts with some earlier studies \citep{1990AJ....100.1489W, 2011MNRAS.418..244M}, which reported a predominance of lenticular and elliptical hosts (likely a consequence of the significantly smaller samples considered in those works). Specifically, the fractions of spiral galaxies among PRs from these catalogs that were included in our catalog are 11.4\% for \citet{1990AJ....100.1489W} and 4.7\% for \citet{2011MNRAS.418..244M}. It was also previously believed that gas-rich PRs could not coexist in a stable configuration with a gas-rich disky host \citep[see e.g.][]{2006ApJ...643..200I}; however, late-type hosts have been noted in the literature for a few PRs \citep{2014ASPC..486...61M}, including the well-known PRG, NGC\,660 \citep{1995AJ....109..942V}. Our results demonstrate unequivocally that late-type galaxies can indeed host PRs, and they constitute the most numerous morphological type in our catalog. Similarly, PHs are found more frequently in spiral hosts than any other host type in our catalog. 

PDLs, which are likely associated with dusty PRs, occur predominantly in elliptical galaxies. As noted above, one explanation for this tendency is likely observational: the rounder shapes and smoother stellar light profiles of ellipticals enhance the visual prominence of central dust absorption features. Both PTSs and extreme warps are also most commonly found in spiral hosts. Strong gravitational perturbations during galaxy interactions and minor mergers can redistribute gas, dust, and stars into complex configurations, including polar streams that may eventually settle into rings, disks, or halos. PBs, on the other hand, are found almost exclusively in spiral galaxies, with only a small fraction in lenticular hosts. The nature of PBs remains uncertain (see Sect.~\ref{sec:reliability}); photometric data alone do not allow for a definitive determination of their physical origin.

While the statistics presented above describe only the fractions of PSG host morphologies within COUGS–DESI, it is also informative to determine the occurrence rate of PSGs across different morphological types in the full SGA. Although we applied our CNN~2 classifier to all galaxies in the SGA, its performance appeared less reliable than that of the automated morphology estimates from \citet{2023MNRAS.526.4768W}. Their approach, developed within the \textit{Galaxy~Zoo~DESI} project, uses deep learning models trained directly on the aggregated votes of thousands of Galaxy~Zoo volunteers. These models capture the complex visual patterns recognized by human classifiers, while maintaining consistency across the large sample of 8.7 million galaxies imaged by DESI Legacy. Because the network was trained on high-quality, consensus-based labels and explicitly designed to reproduce the probability distributions of volunteer responses rather than single deterministic labels, it provides robust, calibrated morphological probabilities over a wide dynamic range in surface brightness, redshift, and galaxy size.

To assign morphological types to all SGA galaxies, we cross-matched the SGA with the Galaxy~Zoo~DESI catalog and analyzed systematic trends between our own robust morphological classifications (for $\sim$20,000 galaxies; see Sect.~2.3) and the automated morphology measurements from \citet{2023MNRAS.526.4768W}. Using this, we quantified how individual Galaxy~Zoo~DESI features (e.g., smoothness, edge-on orientation, spiral arm fraction, bar presence, and bulge prominence) vary across morphological classes. For each feature, we computed distribution differences, Cohen’s~$d$, and the area under the ROC curve (AUC) to identify the most discriminating thresholds. These empirically derived relations were then applied to the entire SGA sample to assign galaxies to broad morphological bins: E, S0, S, and M. This procedure combines the strengths of our visually validated subsample with the statistical completeness and reproducibility of the Galaxy~Zoo~DESI machine-learning measurements, ensuring a uniform and physically motivated morphological classification across all SGA galaxies. The results of this automated classification for all SGA galaxies are provided in the bottom row of Table~\ref{tab:psg_fractions_threeparts}.

Spiral galaxies are the most common type of non-dwarf galaxy in the Local Universe (see Table~\ref{tab:psg_fractions_threeparts} and \citealt{2006MNRAS.373.1389C}) and dominate the overall population of PR hosts in our catalog (Fig.~\ref{fig:hostbarcharts}). However, within the SGA, only about 0.15\% of spiral galaxies exhibit PRs, compared to $\sim$1\% of S0 galaxies and $\sim$0.4\% of ellipticals. The relatively high fraction of PSGs assigned to the M category reflects the ambiguous morphology of these systems where the identification of the primary host galaxy is uncertain. Among PSGs with a confidently identified host, lenticulars are the most frequent host type, with $\sim$2\% of S0 galaxies hosting a polar structure.  

Since both lenticular and elliptical galaxies can form through mergers, it is plausible that some systems currently classified as PTSs --- having reached their present state through merging --- may eventually evolve into stable PSGs. To assess the long-term fate of these systems, and to determine whether any enduring polar structures survive the subsequent relaxation phase in its spatial environment, it is necessary to consider modern cosmological simulations. This analysis will be presented in a forthcoming paper.  

Similarly to the case of PRs, PHs are also far more prevalent in S0 galaxies than in spirals. In contrast, PBs are exceedingly rare across all host types, although spirals remain the most common hosts for PBs within the SGA. The number of PBs (and therefore their inferred fraction in the SGA) is likely underestimated.

\section{Discussion} \label{sec:discussion}

\subsection{Reliability of our classification} \label{sec:reliability}

Our catalog includes PSGs observed over a wide range of inclination angles (see Figs.~\ref{fig:prg_pb_examples} and~\ref{fig:ph_pts_examples} for characteristic examples). Depending on the viewing geometry, polar rings may appear as linear streaks, or exhibit elliptical to nearly circular isophotes in the case of face-on orientations \citep{1990AJ....100.1489W}. In some cases, determining whether a face-on ring is truly polar can be challenging. The inclination of the host galaxy further complicates detection: for example, in face-on disk galaxies, polar structures are especially difficult to identify if they are smaller in extent than the host. In such cases, the polar component may not produce a noticeable excess of light along a limited range of position angles (on both sides of the galaxy center) beyond the circular isophotes of the disk. Furthermore, kinematic analysis of face-on PRs or PSGs with face-on hosts is particularly challenging, since the rotation of both the ring and the host occurs primarily in the plane of the sky, perpendicular to the line of sight. Fully characterizing the morphology and dynamics of such systems remains an interesting challenge for future work. 

In our catalog, we include face-on rings when their visual morphology strongly suggests a polar origin, based on the relative orientation and appearance of the ring and host. This assessment is typically guided by the projected geometry: if the central flattened component extends beyond the boundaries of the ring --- such that one end appears behind the ring and the other in front --- it is likely an edge-on disk galaxy rather than a bar in a face-on normal ringed galaxy. In addition, we have identified a robust sample of more than 2,000 galaxies with face-on rings, many of which are likely to be genuine face-on PRs. These will be analyzed in detail in a forthcoming study.

Another area where morphological classification may be ambiguous is in the case of PBs. These structures are generally small and faint compared to their host galaxies, making it difficult to determine their true nature. The polar components in some PB systems may be unresolved or compact polar rings or disks, as suggested by their comparatively lower apparent magnitudes and smaller optical diameters relative to other PSGs. Meanwhile, some other so-called PBs could represent an intermediate class of polar structures, bridging the scale between inner polar disks (typically a few hundred parsecs in size; \citealt{2003A&A...408..873C,2012AstBu..67..147M}) --- which are not included in our catalog, as they can only be identified in the nearest DESI Legacy galaxies --- and the well-known large PRs whose sizes are comparable to those of their hosts. An alternative explanation is that some PBs may actually be end-on bars misidentified as polar features. In any case, detailed two-dimensional photometric decomposition and spectroscopic follow-up are required for each individual system to determine the true nature of the polar component.

Finally, the distinction between PRs and PHs is not always clear, since PHs appear to have a smooth distribution but may in fact be highly diffuse rings or unresolved tidal structures \citep{2024A&A...681L..15M}. We classify as PHs in our catalog those galaxies whose polar structures exhibit a smooth, oval-shaped light distribution. Higher-resolution photometry as well as spectroscopic analysis is necessary to disentangle the true nature of PHs. For PDLs and PTSs, morphological classification is comparatively straightforward, and the likelihood of confusing them with other PS subtypes is lower.

We do not include major mergers in our catalog explicitly, although some galaxies whose hosts are classified in the M category may be ongoing major mergers. We selected only those mergers for our catalog which exhibit well-defined orthogonal structures, and excluded systems where a mixture of stellar components produces a peculiar morphology with multiple tidal tails.

\subsection{Projection effects}\label{subsec:proj_effects}

\begin{figure}
    \centering
\includegraphics[width=\linewidth]{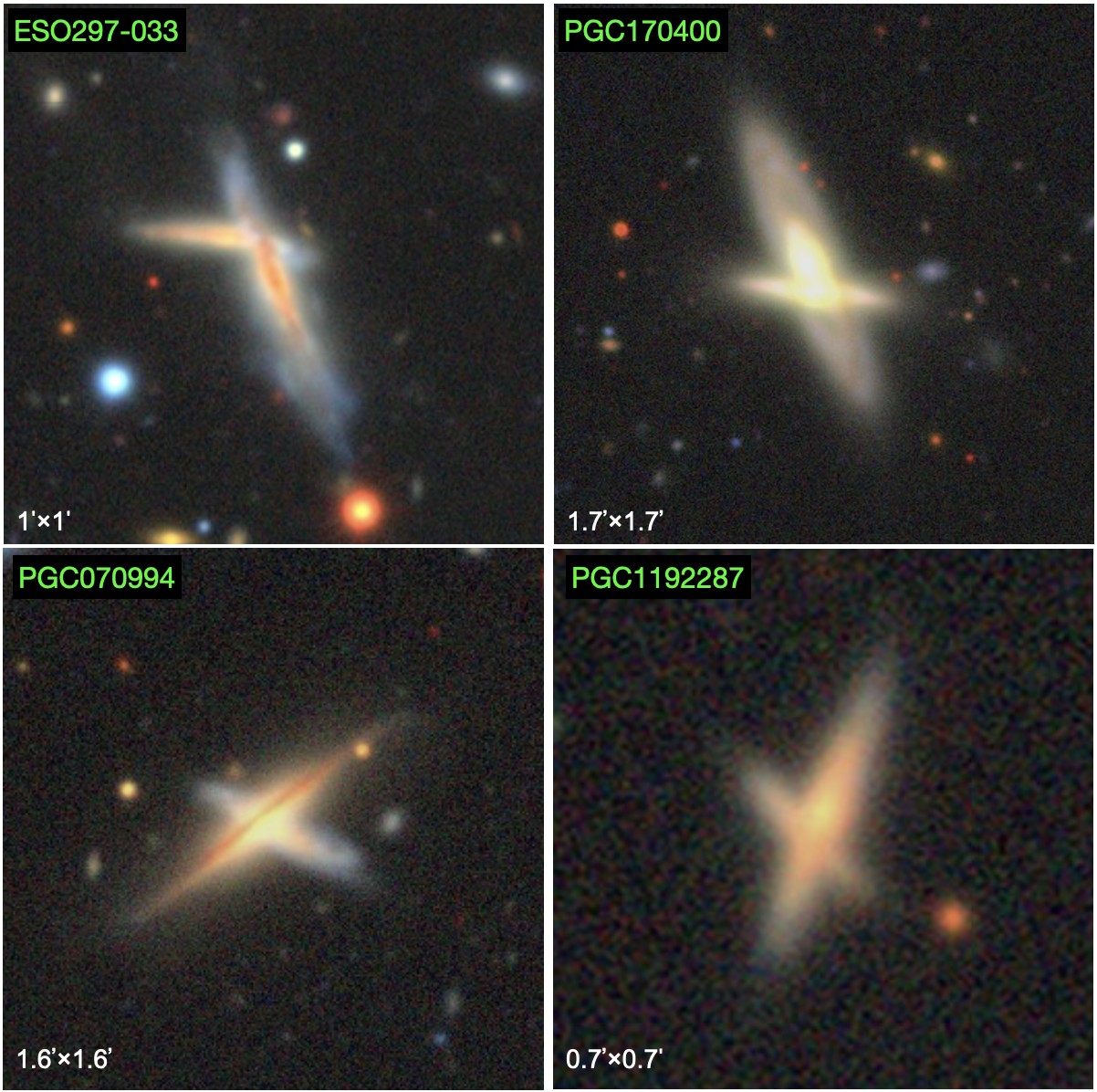}
        \label{fig:topleft}
\caption{Examples of galaxies that appear to host a polar structure due to projection effects, but are likely not genuine PSGs. Careful inspection shows that the smaller galaxies are either background or foreground objects. In contrast, in PSGs from our catalog, the polar structures typically overlap the main galaxy body, passing both in front of and behind it along the line of sight (often evident through dust attenuation by the polar structure).}
    \label{fig:falsepsgs}
\end{figure}

From our vantage point, some objects may appear as PSGs when, in fact, they are a superposition of two unrelated systems along the line of sight, producing an apparent cross-shaped morphology or a false PSG. Examples illustrating this effect are shown in Fig.~\ref{fig:falsepsgs}. Since such line-of-sight overlaps are not always readily identifiable, particularly for faint or compact galaxies, it is essential to quantify how frequently these projection-induced false positives occur to accurately assess the level of contamination in our catalog and, ultimately, to determine the true occurrence rate of PSGs.

To estimate the likelihood of such projection effects, we selected 1000 random fields, each one square degree in size, from the DESI Legacy database. For each field, we measured the number and angular sizes of all detected galaxies, and then randomized their sky positions 1000 times per field, resulting in a total of $10^6$ simulated realizations. To identify cases where galaxies overlapped in projection and could mimic a polar structure, we employed the \texttt{Shapely} Python package to compute intersections between the simulated galaxy footprints.

\begin{figure}
\centering
\includegraphics[width=\linewidth]{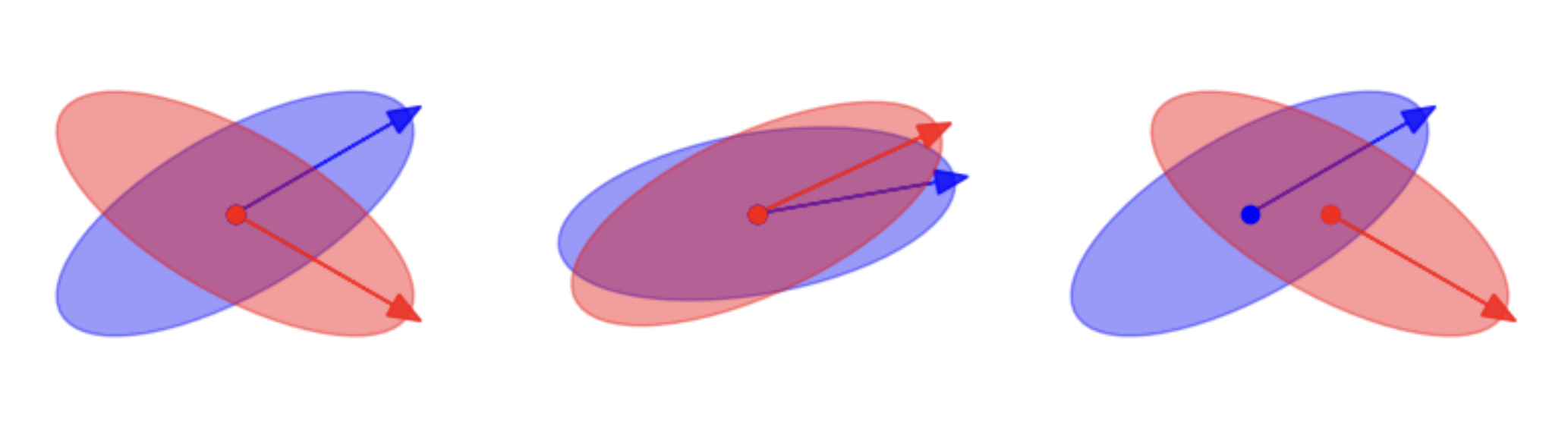}
\caption{
{\it Left:} Ellipses satisfying all three criteria (four intersections, close centers, and polar angles).
{\it Middle:} Ellipses with four intersections and close centers, but not at polar angles.
{\it Right:} Ellipses with four intersections and polar angles, but not close centers.
}
\label{fig:exampleellipses}
\end{figure}

For each simulation, we recorded the number of overlapping galaxies according to several progressively restrictive criteria. First, we identified galaxy ellipses that intersected at four points, creating a cross-shaped morphology. Among these, we then selected pairs with relative position angles of $90\degree \pm 50\degree$, corresponding to a polar orientation. Finally, from this subset, we retained only those overlaps in which the galaxy centers were closely aligned, defined as an offset of less than or equal to one-tenth of the semi-major axis length of the larger galaxy. Fig.~\ref{fig:exampleellipses} illustrates examples of such configurations, showing ellipses with four intersection points, a range of position angles, and varying center separations. Only overlaps satisfying all three criteria (four intersections, close centers, and polar angles) were considered potential false PSGs.

Our simulations identified 6,034 galaxy pairs whose ellipses intersected at four points, corresponding to a probability density of $6.034\times10^{-3}$ per square degree. Normalizing by the average number of galaxies across all 1000 simulated fields (39.527) yields a probability of $1.53\times10^{-4}$ per galaxy. Among these, 4,706 pairs also satisfied the relative position-angle criterion of $90\degree \pm 50\degree$, corresponding to a normalized probability of $1.19\times10^{-4}$. Finally, when additionally requiring that the centers of the two galaxies be separated by no more than one-tenth of the semi-major axis of the larger galaxy, the resulting probability was $2.28\times10^{-5}$ per galaxy. These cases were classified as potential false-positive PSGs.

To estimate the number of false-positive PSGs in our selection due to galaxy overlaps, we multiplied the final probability by the total number of galaxies across the 1000 simulated fields and then by a factor of 20 to scale to the $\sim$20,000~deg$^2$ area of the SGA footprint. This yields an expectation of $\approx18$ galaxies, representing a lower-limit estimate of line-of-sight overlaps that could mimic PSGs. When relaxing only the central-offset criterion, we obtain an expected number of $\approx94$ galaxies with polar-angle overlaps. The results from our visual classification --- during which we flagged 38 galaxies suspected of being overlapping systems (and therefore excluded them from the catalog) --- fall within this range, suggesting that our inspection procedure was largely effective in identifying most false-positive overlaps. Nevertheless, a small number of PSGs in our catalog may ultimately prove to be chance superpositions once kinematic data become available.

\subsection{Occurrence rate of PSGs} \label{sec:oc_rate}

\begin{figure}
    \centering
    \includegraphics[width=\linewidth]{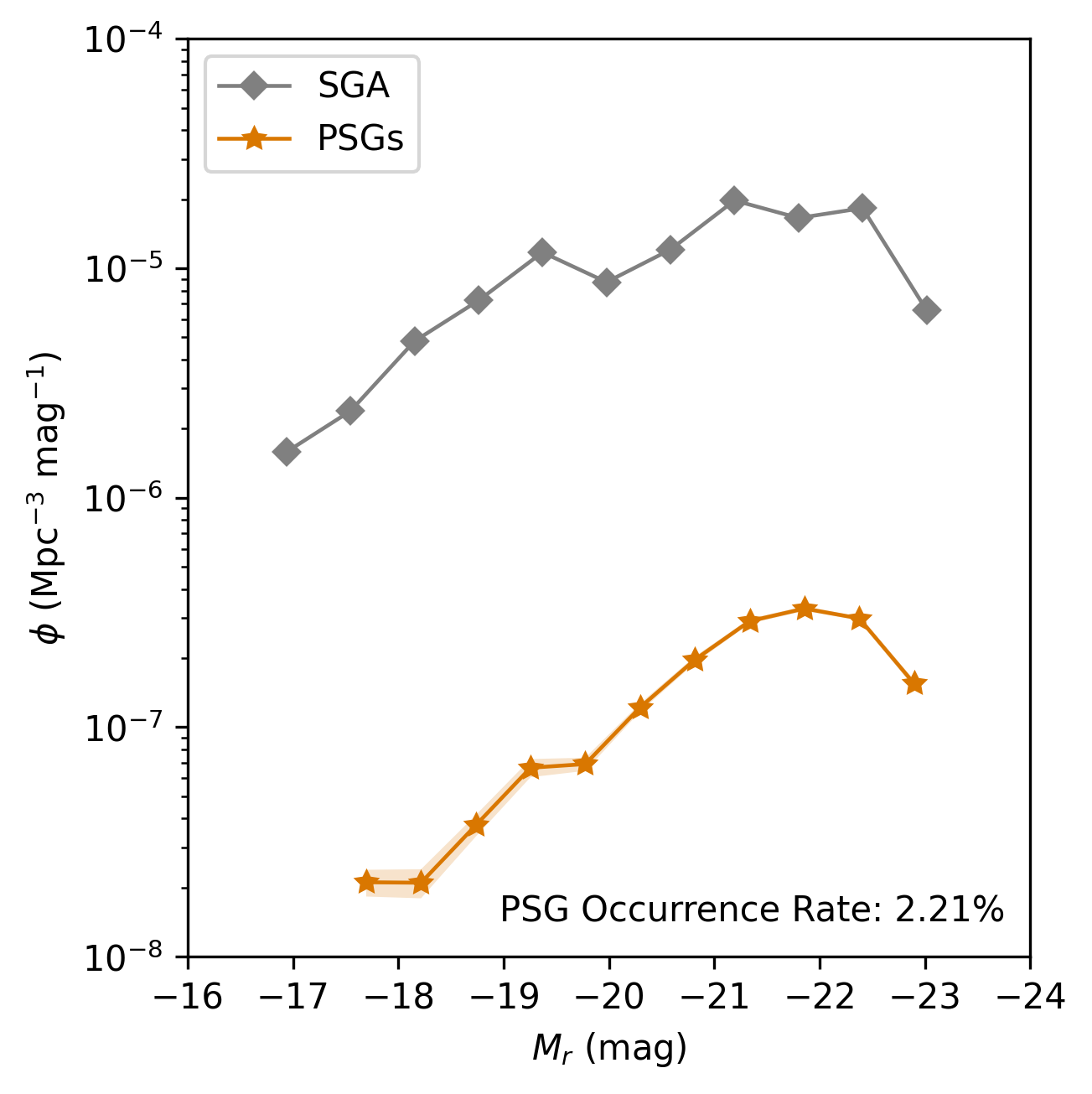}
    \caption{Luminosity functions for the SGA and PSG samples, calculated following the Cho\l oniewski method \citep{1986MNRAS.223....1C}. The curves show the differential space density of galaxies $\phi(M)$ as a function of absolute magnitude $M_r$, with shaded regions indicating uncertainties. The PSG occurrence rate of 2.2\% was derived from the ratio of the integrated luminosity functions, each corrected using the same methodology.}
    \label{fig:full_LF}
\end{figure}

\begin{figure}
    \centering
    \includegraphics[width=\linewidth]{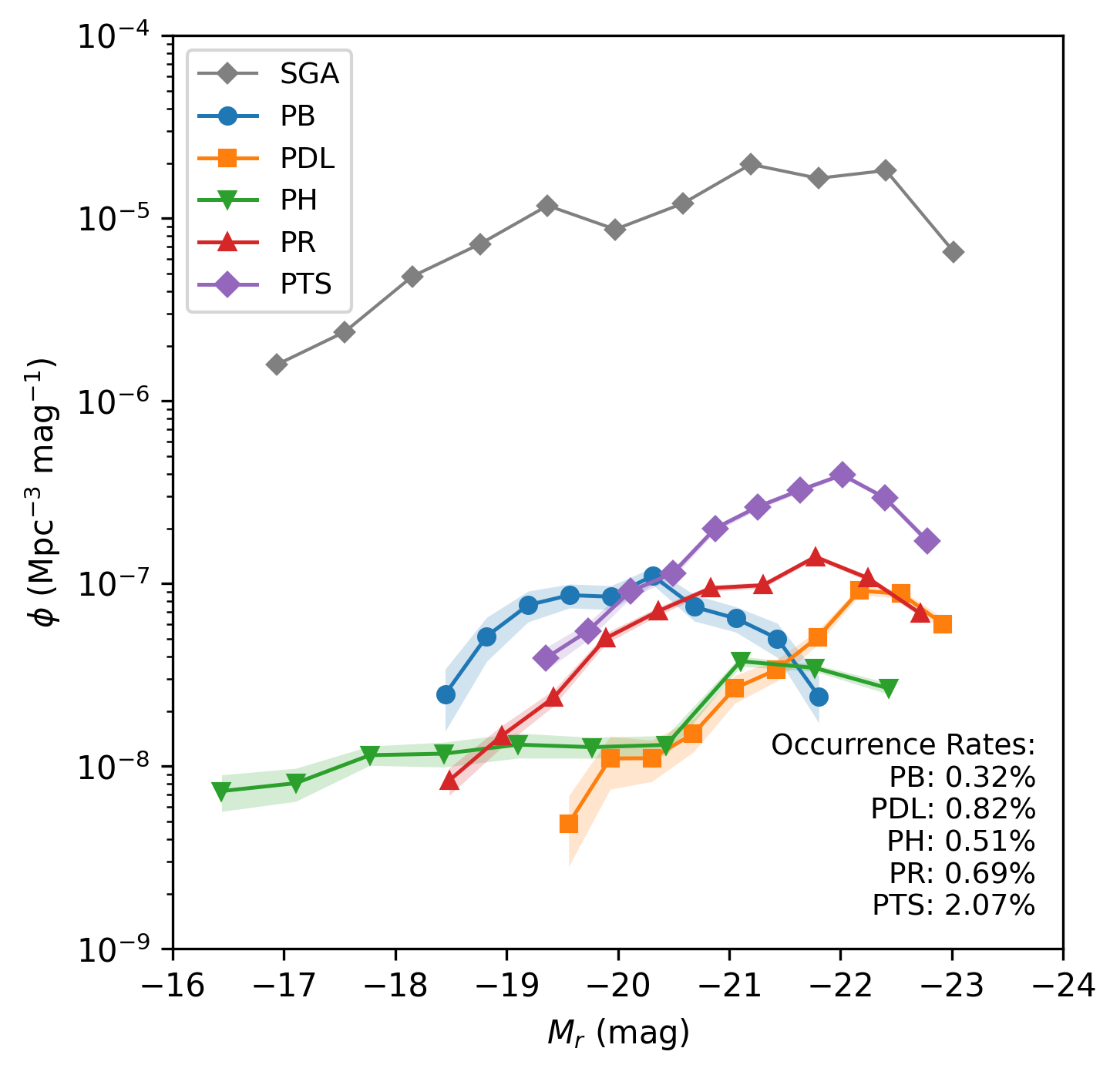}
    \caption{Same as Fig.~\ref{fig:full_LF} but for different subtypes of PSGs. Occurrence rates for each subtype were derived from the ratio of their integrated luminosity functions to that of the SGA, each corrected using the same methodology.}
    \label{fig:subtype_LF}
\end{figure}

To estimate the occurrence rate of PSGs, we calculated the luminosity function for our sample using the Cho\l oniewski method \citep{1986MNRAS.223....1C}, as demonstrated by \citet{2022MNRAS.516.3692S}. This non-parametrical method is designed specifically for calculating the luminosity function of a magnitude-limited sample where the homogeneous distribution of galaxies is not assumed. Unlike other non-parametrical methods, the Cho\l oniewski method provides a normalized luminosity function with errors. We applied this method in its original form, except for changing the integration method to a Simpson method to avoid double-counting. The luminosity function for the entire sample of PSGs is shown in Fig.~\ref{fig:full_LF}. As in Fig.~\ref{fig:gen_stats_zoomed}, PSGs are slightly overrepresented at bright absolute magnitudes compared to the entire SGA sample. In addition, the luminosity function for each individual PSG subtype, calculated by the same method, is shown in Fig.~\ref{fig:subtype_LF}. 

Inspecting each PSG subtype individually, we see that the differential space densities of PDLs and PTSs peak at $M_r\approx-22$ and decrease rapidly for dimmer galaxies. These structures are quite faint and difficult to identify, even more so in faint host galaxies. A similar pattern exists for PRs, but the drop as a function of magnitude is slightly less pronounced, and the peak of the luminosity function is slightly shifted toward dimmer magnitudes. Meanwhile, the luminosity function for PB galaxies peaks at $M_r\approx-20.3$ and falls off in both directions. This matches the typically lower luminosity of PB galaxies compared to other PSG subtypes, as discussed in Sect.~\ref{subsec:stat}. Finally, the luminosity function for PHs behaves very differently. It is almost flat from $M_r=-16$ to $-20$ and $-21$ to $-23$, but with a large jump between $M_r=-20$ and $-21$.

We derived the PSG occurrence rate from the ratio of the integrated luminosity functions for the entire SGA and the PSG sample. We found a PSG occurrence rate of 2.2\% of local non-dwarf galaxies, in good agreement with the previous estimate of 1--3\% by \citet{2024A&A...681L..15M}. Breaking down the luminosity function for PSGs into the different subtypes, we see that the subtype with the largest contribution to the occurrence rate is PTSs (2.1\%). The next highest contribution is by PDLs (0.8\%), followed by PRs (0.7\%). PHs and PBs are rarer, with occurrence rates of 0.5\% and 0.3\%, respectively.

While this catalog contains a much larger number of PSGs than previously known, we anticipate that there are still a few PSGs missing from the catalog. Polar structures that are very dim, diffuse, or primarily gaseous are difficult to identify through the optical photometry currently available. In most cases, polar structures are more easily identifiable in highly inclined galaxies than in face-on galaxies. As a result, our catalog is dominated by highly inclined galaxies. The influence of this inclination effect on the occurrence rate of PSGs, based on ellipse fitting of the galaxies in our catalog, will be determined in a subsequent paper. For even more accurate measurements of the luminosity function and occurrence rate of PSGs in the Local Universe, we will need even deeper observations with better angular resolution. The Euclid telescope is anticipated to provide the necessary data, and an expanded analysis using Euclid's Quick Data Release 1 \citep[][submitted to A\&A]{2025arXiv250315302E} will be presented in a forthcoming paper.

\subsection{Future studies} \label{subsec:future}

COUGS-DESI and the accompanying atlas open pathways for a wide variety of future investigations. As the largest sample of PSGs compiled to date, COUGS-DESI not only offers a rich set of intriguing individual systems for detailed study, but also enables robust statistical analyses aimed at improving our understanding of the PSG population as a whole.

The diversity of PSG subtypes identified in our catalog indicates that further investigation of their structural properties is essential for a comprehensive characterization of these systems. Our classification relies on visual morphology derived from optical photometry. However, certain polar structures, particularly PBs and PHs, can appear ambiguous in optical images. In such cases, photometric decomposition provides a valuable tool for disentangling the components and clarifying the true nature of the polar structures. Such methods may also clarify our definitions of the PSG subtypes. The effectiveness of this approach for PSGs is demonstrated in our pilot study by \citet{2025A&A...698L..21B}.

Since PSGs are believed to originate primarily through galaxy interactions or the accretion of cold gas from cosmological filaments, many are expected to exhibit LSB features. Deep photometric observations enable the study of these faint structures, offering insights into the formation and evolution of polar structures through mergers, tidal interactions, and cold accretion. Moreover, examining the spatial environments of PSGs will provide additional clues about the mechanisms that drive these processes.

Another key step toward understanding the formation of PSGs involves cosmological simulations. Such simulations allow the reconstruction of merger trees for individual galaxies, thereby revealing possible formation pathways for PSGs, as demonstrated in the pilot study by \citet{smirnov2024}. We have already identified a sample of PSGs within the Illustris~TNG50 simulation \citep{2019MNRAS.490.3196P,2019MNRAS.490.3234N}, and a detailed comparison between these simulated systems and the observational sample presented in this catalog will be provided in a forthcoming paper. Future work will also investigate the formation histories of simulated PSGs in greater depth, including comparing formation mechanisms of various PSG subtypes in detail. 

In addition, surveys beyond DESI Legacy offer valuable opportunities to advance our understanding of PSG evolution. Infrared observations, in particular, help mitigate the obscuring effects of dust lanes that are common in the optical DESI Legacy data. Ongoing and planned efforts aim to identify PSGs in deep extragalactic surveys conducted with the \textit{James Webb Space Telescope} \citep{2023PASP..135f8001G}, the \textit{Euclid} mission \citep{2025A&A...697A...1E}, and the forthcoming \textit{Nancy Grace Roman Space Telescope} (originally the \textit{Wide-Field Infrared Survey Telescope}; \citealt{2015arXiv150303757S}). These datasets will enable the extension of PSG studies across a broader range of wavelengths and redshifts, providing critical constraints on the formation and evolution of PSGs.

Many galaxies in our catalog exhibit warps, lopsidedness, and flaring in both host components and polar structures. These morphological distortions are likely signatures of external accretion or gravitational perturbations. In particular, the rings of PRGs are frequently warped due to the combined effects of differential precession, the triaxial shape of the host galaxy’s gravitational potential, and external tidal interactions (see e.g., \citealt{2006EAS....20...97C} and references therein). Misalignments between the angular momenta of the host and the accreted material can also induce long-lived warps or tilts in disk galaxies \citep{2010MNRAS.408..783R}. Simulations have shown that warps in PRs may persist for several gigayears if the dark matter halo is sufficiently flattened or misaligned with the stellar disk \citep{2003A&A...401..817B}. Additional photometric analysis of such systems is required to determine how these features originate and how they influence the longevity, star formation rate, and overall evolution of PSGs.

A subsample of relatively large PSGs in our catalog is well suited for follow-up slit and integral-field unit (IFU) spectroscopy, which can be used to investigate the kinematics and dynamics of both the host galaxy and the polar structure. Such observations will enable detailed kinematic decomposition, allowing the reconstruction of the intrinsic three-dimensional shapes of dark matter halos in PSGs (see, e.g., \citealt{2014MNRAS.441.2650K}). Combined stellar and gaseous velocity fields will also help constrain the degree of orthogonality between the host and polar components, revealing the dynamical stability of these systems and the role of angular momentum transfer during their formation. Furthermore, IFU data can provide spatially resolved information on star formation rates, metallicity gradients, and ionization conditions across the polar and host components, offering a comprehensive view of how gas accretion, mergers, and secular evolution shape the observed diversity of PSGs.

The star formation and gas dynamics of PRs in early-type galaxies (ETGs) differ markedly from those of typical ETGs \citep{2015MNRAS.447.2287R,2019MNRAS.486.4186E,2024Galax..12...42L}. Most PRGs show enhanced star formation within their polar rings, which gives rise to their characteristic blue colors \citep{1999ASPC..163..197E}. Moreover, many PRGs host active galactic nuclei \citep{2020AstL...46..501S}, although it remains unclear whether this trend extends to other PSG subtypes. Future studies should examine the roles of star formation as well as AGN activity and feedback in shaping the evolution of PSGs.

\section{Summary and conclusions} \label{sec:summary}

We present COUGS-DESI, the largest catalog and atlas of polar-structure galaxies (PSGs) to date, comprising 2,989 objects primarily drawn from the Siena Galaxy Atlas \citep{2023ApJS..269....3M} based on the DESI Legacy Imaging Surveys \citep{2019AJ....157..168D} DR10. The catalog includes galaxies exhibiting a wide range of polar structures: 1,113 polar-ring (PR) galaxies, 75 polar-bulge (PB) galaxies, 216 polar-halo (PH) galaxies, 185 galaxies with polar dust lanes (PDLs), and 1,315 galaxies with polar tidal structures (PTSs). It also contains an incomplete sample of related objects with extreme warps and polar X-shaped bulges.

The catalog was compiled through a combination of complementary approaches: (1) reviewing existing catalogs and published reports of PSGs in the literature; (2) developing CNNs to automatically classify galaxy images; and (3) performing a manual search for PSGs in DESI Legacy~DR10 photometry based on the \textit{Siena Galaxy Atlas} \citep{2023ApJS..269....3M}. Candidate PSGs identified by these methods were cross-matched to remove duplicates, false positives were excluded, and the remaining systems were consolidated into a comprehensive catalog of PSGs detected in the DESI Legacy survey, together with relevant information about their general properties and morphologies. In addition, we generated enhanced composite images of these galaxies, which are presented in an accompanying atlas. 

In this paper, we also present an initial statistical analysis of the sample of PSG candidates, including their apparent and absolute magnitude, color redshift, diameter, host type, luminosity function, and occurrence rate compared to the full sample of galaxies from the SGA. The main conclusions of our study are as follows:

\begin{itemize}
    \item Although we have refined the nomenclature for visually classifying PSGs into distinct subtypes, the boundaries between these categories remain somewhat subjective and can be ambiguous when based solely on visual inspections.
    \item Both red and blue polar structures are observed in both red and blue host galaxies (see examples in Figs.~\ref{fig:prg_pb_examples}--\ref{fig:ph_pts_examples}), indicating no strict color dependence between the two components.
    \item Simple image simulations of DESI Legacy fields indicate that projection effects from randomly overlapping galaxies, which could otherwise produce false-positive PSG candidates, are minimal and do not introduce any contamination in our catalog.
    \item Compared to the general SGA galaxy population, PSGs have, on average, brighter apparent magnitudes (but comparable absolute magnitudes), slightly lower redshifts, larger angular sizes, and redder optical colors (Fig.~\ref{fig:gen_stats_zoomed}).
    \item The fraction of identifiable PSGs in DESI Legacy images decreases significantly with increasing redshift beyond $z\gtrsim0.05$ (Fig.~\ref{fig:gen_stats_zoomed}) due to selection effects: at higher redshifts, polar structures become fainter and less distinct as a result of decreasing spatial resolution. 
    \item Among PSG subtypes, PB galaxies tend to have fainter apparent and absolute magnitudes, smaller angular sizes, and bluer optical colors compared to other PSGs (Fig.~\ref{fig:gen_stats_zoomed_subtypes}). PBs are predominantly associated with spiral hosts (Fig.~\ref{fig:hostbarcharts}).
    \item PDL galaxies, which predominantly occur in elliptical hosts (Fig.~\ref{fig:hostbarcharts}), exhibit brighter apparent and absolute magnitudes, larger angular sizes, and redder optical colors than other PSG subtypes (Fig.~\ref{fig:gen_stats_zoomed_subtypes}).
    \item In contrast to previous studies of PSGs, we observe that for PRs, spirals are the most common host type in our catalog (Fig.~\ref{fig:hostbarcharts}), indicating that the fraction of polar rings in late-type galaxies is higher than previously thought.
    \item Among host galaxies with clearly identifiable morphologies, lenticular galaxies show the highest fraction of polar structures for most PSG subtypes. Approximately 1\% of S0 galaxies in the SGA exhibit polar rings (Table~\ref{tab:psg_fractions_threeparts}).
    \item Based on the luminosity function derived from our catalog, PSGs constitute approximately 2.2\% of local non-dwarf galaxies (Fig.~\ref{fig:full_LF}).
    \item When excluding PTSs, PR galaxies are the most common PSG subtype, with an occurrence rate of about 0.7\% (Fig.~\ref{fig:subtype_LF}).
\end{itemize}

The new catalog and atlas of PSGs, COUGS–DESI, represents a valuable resource for future studies of these unique objects. Because PSGs serve as excellent laboratories for exploring the physical processes that govern galaxy formation and evolution, this significantly expanded sample will enable more comprehensive investigations in these areas. A series of forthcoming papers based on COUGS–DESI will address multiple aspects of PSG research, including detailed photometric decomposition, refined morphological classification, the detection and characterization of LSB features in PSGs, analyses of spatial environments, comparisons with cosmological simulations, and studies of PSG kinematics, dynamics, star formation, and AGN activity.

\section*{Data availability}
The catalog is only available in electronic form at the CDS via anonymous ftp to cdsarc.u-strasbg.fr (130.79.128.5) or via http://cdsweb.u-strasbg.fr/cgi-bin/qcat?J/A+A/710/A145. The atlas is only available in electronic form at Zenodo via https://zenodo.org/records/20042619.

\begin{acknowledgements}

We thank the anonymous referee for their helpful comments.

We acknowledge the help of the following students during the classification process: 
Emma Aguirre,
James Bleazard,
Steven Blodgett,
JoAnn Castellon,
Perri Coggins,
Scott Curtis,
Colin Derieg,
Nathanael Garey,
Carter Garrett,
Megan Gee,
Peter Jensen,
Tyler Jensen,
Regal Ledbetter,
Brandon Matheson,
Joshua Miller,
Sam Norcross,
Madeline Rotz,
Carson Tenney,
Lydia Stacey,
Megan Tsai,
Nathan Van Dyke,
Kade Vickers, and
Joe Williams.

The Legacy Surveys consist of three individual and complementary projects: the Dark Energy Camera Legacy Survey (DECaLS; NOAO Proposal ID \# 2014B-0404; PIs: David Schlegel and Arjun Dey), the Beijing-Arizona Sky Survey (BASS; NOAO Proposal ID \# 2015A-0801; PIs: Zhou Xu and Xiaohui Fan), and the Mayall z-band Legacy Survey (MzLS; NOAO Proposal ID \# 2016A-0453; PI: Arjun Dey). DECaLS, BASS and MzLS together include data obtained, respectively, at the Blanco telescope, Cerro Tololo Inter-American Observatory, National Optical Astronomy Observatory (NOAO); the Bok telescope, Steward Observatory, University of Arizona; and the Mayall telescope, Kitt Peak National Observatory, NOAO. The Legacy Surveys project is honored to be permitted to conduct astronomical research on Iolkam Du’ag (Kitt Peak), a mountain with particular significance to the Tohono O’odham Nation.

We acknowledge the usage of the HyperLeda database (http://leda.univ-lyon1.fr).

IRAF is distributed by the National Optical Astronomy Observatory, which is operated by the Association of Universities for Research in Astronomy (AURA) under cooperative agreement with the National Science Foundation \citep{1993ASPC...52..173T}.

This research has made use of the Astrophysics Data System, funded by NASA under Cooperative Agreement 80NSSC25M7105.

This research has made use of the NASA/IPAC Extragalactic Database (NED), which is funded by the National Aeronautics and Space Administration and operated by the California Institute of Technology.

The Photometric Redshifts for the Legacy Surveys (PRLS) catalog used in this paper was produced thanks to funding from the U.S. Department of Energy Office of Science, Office of High Energy Physics via grant DE-SC0007914.

Funding for the Sloan Digital Sky Survey V has been provided by the Alfred P. Sloan Foundation, the Heising-Simons Foundation, the National Science Foundation, and the Participating Institutions. SDSS acknowledges support and resources from the Center for High-Performance Computing at the University of Utah. SDSS telescopes are located at Apache Point Observatory, funded by the Astrophysical Research Consortium and operated by New Mexico State University, and at Las Campanas Observatory, operated by the Carnegie Institution for Science. The SDSS web site is \url{www.sdss.org}.
SDSS is managed by the Astrophysical Research Consortium for the Participating Institutions of the SDSS Collaboration, including the Carnegie Institution for Science, Chilean National Time Allocation Committee (CNTAC) ratified researchers, Caltech, the Gotham Participation Group, Harvard University, Heidelberg University, The Flatiron Institute, The Johns Hopkins University, L'Ecole polytechnique f\'{e}d\'{e}rale de Lausanne (EPFL), Leibniz-Institut f\"{u}r Astrophysik Potsdam (AIP), Max-Planck-Institut f\"{u}r Astronomie (MPIA Heidelberg), Max-Planck-Institut f\"{u}r Extraterrestrische Physik (MPE), Nanjing University, National Astronomical Observatories of China (NAOC), New Mexico State University, The Ohio State University, Pennsylvania State University, Smithsonian Astrophysical Observatory, Space Telescope Science Institute (STScI), the Stellar Astrophysics Participation Group, Universidad Nacional Aut\'{o}noma de M\'{e}xico, University of Arizona, University of Colorado Boulder, University of Illinois at Urbana-Champaign, University of Toronto, University of Utah, University of Virginia, Yale University, and Yunnan University.

The Siena Galaxy Atlas was made possible by funding support from the U.S. Department of Energy, Office of Science, Office of High Energy Physics under Award Number DE-SC0020086 and from the National Science Foundation under grant AST-1616414.

\end{acknowledgements}

\bibliographystyle{aa}
\bibliography{bibliography}

\begin{appendix}

\section{Creating enhanced RGB images} \label{app:rgb_method}

The PSG catalog is accompanied by an atlas that includes an enhanced RGB image for each galaxy. These images are specifically designed to reveal LSB features in the galaxy outskirts while preserving the morphology of the bright central regions. Faint structures in the periphery are often best discerned at higher image brightness levels; however, globally brightening an image typically leads to saturation and loss of detail in the core. 

To make both the inner and outer components simultaneously visible, we developed a dedicated technique, termed isophote-scheduled RGB rendering. This algorithm allows us to preserve high–surface-brightness structures in galaxy interiors while enhancing very LSB outskirts --- achieving a balance that is difficult to obtain with conventional $\mathrm{log}$ or $\mathrm{asinh}$ intensity stretches.

Starting from calibrated $g$, $r$, and $z$ FITS frames, we form an SDSS-style RGB composite with fixed band weights and a mild global $\mathrm{asinh}$ tone mapping to set color ratios. We convert the $r$-band image to a surface-brightness map \[
\mu(\mathbf{x}) \;=\; 22.5 \;-\; 2.5\,\log_{10}\!\left[\frac{I_r(\mathbf{x})}{\mathrm{(0.262\,arcsec/pix)}^2}\right],
\] and discretize $\mu$ into isophote bins between $\mu_{\min}=20.5$~mag\,arcsec$^{-2}$ and $\mu_{\max}=28.5$~mag\,arcsec$^{-2}$ with step $\Delta\mu=0.1$~mag\,arcsec$^{-2}$.

For each bin we assign three display parameters ---bias, contrast, and brightness --- that vary linearly with $\mu$ about an anchor level $\mu_{\rm anchor}$:
\begin{align}
p(\mu) &= p_{\rm anchor}
        + \left(\frac{\mathrm{d}p}{\mathrm{d}\mu}\right)\,(\mu - \mu_{\rm anchor}), \nonumber\\
p &\in \{\mathrm{bias,\,contrast,\,brightness}\}.
\end{align}
with each parameter clipped to predefined bounds to ensure stability.

We convert the RGB image to the hue–saturation–value (HSV) color space and operate on the value channel $V$. For each pixel we apply a monotonic stretch $\mathcal{S}$ (one of \texttt{linear}, \texttt{log}, \texttt{sqrt}, \texttt{squared}, \texttt{asinh}, or a histogram-based variant) using the bin-appropriate bias and contrast:
\[
V' = \mathcal{S}\,\bigl(\mathrm{normalize}(V;\,\mathrm{bias},\mathrm{contrast})\bigr),
\]
then we scale by brightness to obtain the final value
\[
V_{\mathrm{final}} = \mathrm{clip}\,\bigl(V'\times \mathrm{brightness},\,0,\,1\bigr).
\]
We recombine $V_{\mathrm{final}}$ with the original hue and saturation to preserve chroma while adjusting local luminance.

To mitigate noise at low surface brightnesses without degrading inner detail, we apply a Gaussian blur with a width of $\sigma = 1$~pix to regions with surface brightness fainter than 24~mag\,arcsec$^{-2}$ in the $r$ band.

By scheduling bias/contrast/brightness explicitly as functions of surface brightness $\mu$, we ensure that galaxy cores remain unsaturated and detailed, whereas tidal debris, PRs, and other very LSB structures are cleanly lifted above the background --- outperforming uniform $\log$ or $\mathrm{asinh}$ stretches in high dynamic-range scenes.

\begin{figure}
\centering
\includegraphics[width=\linewidth]{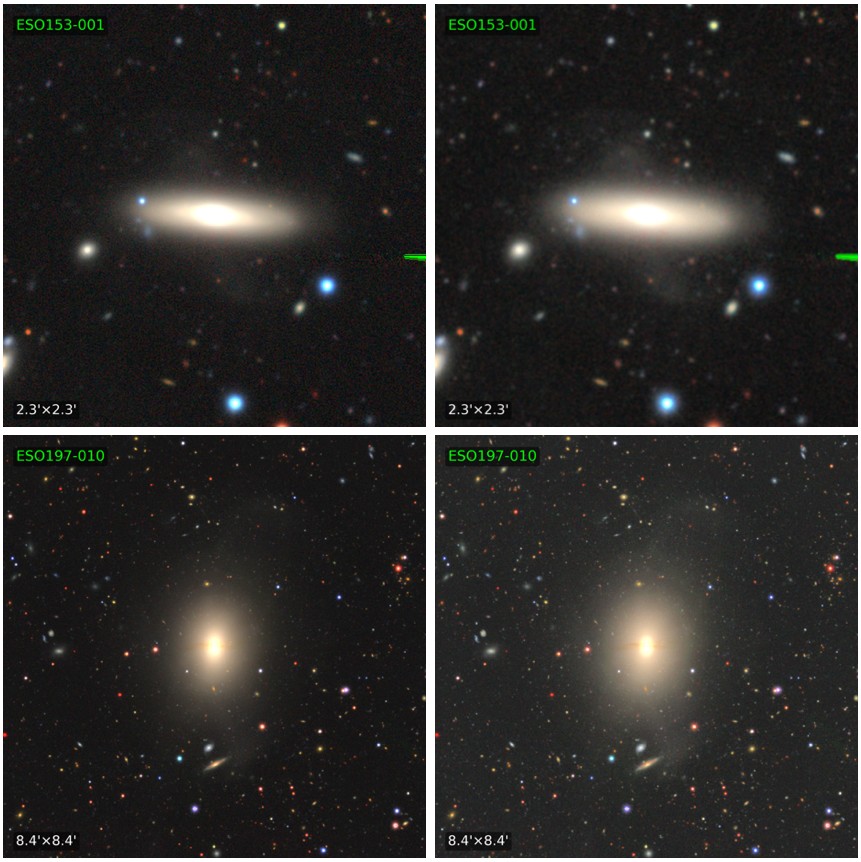}
\caption{Examples of standard SDSS-style RGB images (left column) and enhanced RGB images produced by our algorithm (right column). The enhanced images reveal inner galactic structures and faint outer tidal features more clearly.
}
\label{fig:betterRGB}
\end{figure}

The chosen combination of parameters (bias, contrast, brightness, and $\sigma$) was found by empirically testing many different values for each parameter until all features were more clearly visible and the images were visually appealing. Examples of the enhanced RGB images compared to SDSS-style RGB images are shown in Fig.~\ref{fig:betterRGB}.

\section{The COUGS V1 table} \label{app:table}

\begin{table*}[htbp]
\centering
\caption{Column descriptions for the COUGS v1 catalog.}
\begin{tabular}{lll}
\hline
Column Name & Units & Description \\
\hline
NAME &  & Galaxy name in the COUGS v1 catalog. \\
IAU\_ID &  & IAU-designated galaxy identifier based on RA and DEC. \\
RA & deg & Right Ascension (J2000, decimal degrees) of the galaxy center. \\
DEC & deg & Declination (J2000, decimal degrees) of the galaxy center. \\
PSG\_TYPE\_1 &  & Primary structural classification of the PSG. \\
PSG\_TYPE\_2 &  & Secondary structural classification of the PSG. \\
MORPHTYPE &  & Host galaxy morphological type (E, S0, S, Irr, M). \\
PGC &  & Principal Galaxy Catalog identifier. \\
MORPHTYPE\_LEDA &  & Morphological classification from LEDA. \\
SGA\_ID &  & ID number from the SGA catalog. \\
G\_MAG\_SB26\_IRAF & mag & g-band magnitude measured at 26 mag\,arcsec$^{-2}$ level (IRAF). \\
G\_MAG\_SB26\_ERR\_IRAF & mag & Uncertainty of the g-band asymptotic magnitude. \\
R26\_G\_IRAF & arcmin & g-band isophotal radius at 26 mag\,arcsec$^{-2}$. \\
R26\_G\_ERR\_IRAF & arcmin & Uncertainty of the g-band isophotal radius at 26 mag\,arcsec$^{-2}$. \\
R\_MAG\_SB26\_IRAF & mag & r-band magnitude measured at 26 mag\,arcsec$^{-2}$ level (IRAF). \\
R\_MAG\_SB26\_ERR\_IRAF & mag & Uncertainty of the r-band asymptotic magnitude. \\
R26\_R\_IRAF & arcmin & r-band isophotal radius at 26 mag\,arcsec$^{-2}$. \\
R26\_R\_ERR\_IRAF & arcmin & Uncertainty of the r-band isophotal radius at 26 mag\,arcsec$^{-2}$. \\
I\_MAG\_SB26\_IRAF & mag & i-band magnitude measured at 26 mag\,arcsec$^{-2}$ level (IRAF). \\
I\_MAG\_SB26\_ERR\_IRAF & mag & Uncertainty of the i-band asymptotic magnitude. \\
R26\_I\_IRAF & arcmin & i-band isophotal radius at 26 mag\,arcsec$^{-2}$. \\
R26\_I\_ERR\_IRAF & arcmin & Uncertainty of the i-band isophotal radius at 26 mag\,arcsec$^{-2}$. \\
Z\_MAG\_SB26\_IRAF & mag & z-band magnitude measured at 26 mag\,arcsec$^{-2}$ level (IRAF). \\
Z\_MAG\_SB26\_ERR\_IRAF & mag & Uncertainty of the z-band asymptotic magnitude. \\
R26\_Z\_IRAF & arcmin & z-band isophotal radius at 26 mag\,arcsec$^{-2}$. \\
R26\_Z\_ERR\_IRAF & arcmin & Uncertainty of the z-band isophotal radius at 26 mag\,arcsec$^{-2}$. \\
A\_G & mag & Galactic extinction in g band based on \citet{2011ApJ...737..103S}. \\
A\_R & mag & Galactic extinction in r band based on \citet{2011ApJ...737..103S}. \\
A\_I & mag & Galactic extinction in i band based on \citet{2011ApJ...737..103S}. \\
A\_Z & mag & Galactic extinction in z band based on \citet{2011ApJ...737..103S}. \\
REDSHIFT &  & Final adopted redshift \\
REDSHIFT\_SOURCE &  & Name of the column that supplied REDSHIFT for that row \\
DA & Mpc & Angular diameter distance. \\
SCALE & kpc/arcsec & Physical scale. \\
DL & Mpc & Luminosity distance. \\
KCOR\_G & mag & K-corrections in g band using K-corrections calculator. \\
KCOR\_R & mag & K-corrections in r band using K-corrections calculator. \\
KCOR\_I & mag & K-corrections in i band using K-corrections calculator. \\
KCOR\_Z & mag & K-corrections in z band using K-corrections calculator. \\
M\_G & mag & Absolute magnitude in g band corrected for Galactic extinction and K-correction. \\
M\_R & mag & Absolute magnitude in r band corrected for Galactic extinction and K-correction. \\
M\_I & mag & Absolute magnitude in i band corrected for Galactic extinction and K-correction. \\
M\_Z & mag & Absolute magnitude in z band corrected for Galactic extinction and K-correction. \\
PHYS\_SMA\_SB26 & kpc & Physical semi-major axis at the r-band SB26 level - computed using R26\_R\_IRAF. \\
\hline
\end{tabular}
\end{table*}

\end{appendix}

\label{LastPage}

\end{document}